\documentclass[fleqn,usenatbib]{mnras}

\usepackage{newtxtext,newtxmath}

\usepackage[T1]{fontenc}

\DeclareRobustCommand{\VAN}[3]{#2}
\let\VANthebibliography\thebibliography
\def\thebibliography{\DeclareRobustCommand{\VAN}[3]{##3}\VANthebibliography}

\usepackage{graphicx}   
\usepackage{amsmath}    
\usepackage{pdflscape}  
\usepackage{xcolor}             

\title[Differentially rotating equilibrium models]{Equilibrium sequences of differentially rotating stars with post-merger-like rotational profiles}

\author[P. Iosif and N. Stergioulas]{
Panagiotis Iosif$^{1}$\thanks{E-mail: piosif@auth.gr}
and Nikolaos Stergioulas$^{1}$
\\
$^{1}$Department of Physics, Aristotle University of Thessaloniki, Thessaloniki 54124, Greece
}

\date{Accepted XXX. Received YYY; in original form ZZZ}

\pubyear{2021}

\begin{document}

\label{firstpage}
\pagerange{\pageref{firstpage}--\pageref{lastpage}}
\maketitle

\begin{abstract}
We present equilibrium sequences of rotating relativistic stars, constructed with a new rotation law that was proposed by Uryu et al. (2017). We choose rotational parameters motivated by simulations of binary neutron star merger remnants, but otherwise adopt a cold, relativistic $N=1$ polytropic EOS, in order to perform a detailed comparison to published equilibrium sequences that used the Komatsu, Eriguchi and Hachisu (1989) rotation law. We find a small influence of the choice of rotation law on the mass of the equilibrium models and a somewhat larger influence on their radius. The versatility of the new rotation law allows us to construct models that have a similar rotational profile and axis ratio as observed for merger remnants, while at the same time being quasi-spherical. More specifically, we construct equilibrium sequence variations with different degrees of differential rotation and identify type A and type C solutions, similar to the corresponding types in the classification of Ansorg, Gondek-Rosi{\'n}ska and Villain (2009). While our models are highly accurate solutions of the fully general relativistic structure equations, we demonstrate that for models relevant to merger remnants  the IWM-CFC approximation still maintains an acceptable accuracy. 
\end{abstract}




\section{Introduction}
\label{sec:intro}

Gravitational waves detected from the GW170817 neutron star merger event \citep{GW170817_observation} and complementary information from its  electromagnetic counterpart \citep{GW170817_EM_counterpart, Goldstein_etal_2017} were pivotal in confirming the broad picture describing the possible outcomes of a binary neutron star (BNS) merger. The behaviour of the binary is closely tied to the equation of state (EOS) and the total mass $M=m_1 + m_2$ of the system, where $m_1$ and $m_2$ are the binary's components masses. Depending on where exactly $M$ stands compared to three distinct fundamental mass values, will determine the evolution of the BNS (see \citet{Shibata_Hotokezaka_2019, Bernuzzi_2020, Friedman_Stergioulas_2020} for recent reviews). 

A first checkpoint is the comparison of the total mass $M$ to the $M_\text{thres}$ value, which is the threshold mass above which the merger remnant immediately collapses to a black hole. If $M<M_\text{thres}$, the merger results in a hot, massive and differentially rotating, compact object with a substantial material disk around it. In this scenario, another mass comes into play, namely the maximum mass of a cold, uniformly rotating neutron star, $M_\text{max,rot}$. If $M>M_\text{max,rot}$, the remnant can survive several tens of milliseconds (ms) due to the support of differential rotation and thermal pressure. Collapse to a black hole is inevitable though. Dissipative effects (e.g. effective viscosity due to the development of the magneto-rotational instability, see  \citet{Shibata_Hotokezaka_2019, Ciolfi_2020} for recent reviews and also \citet{Radice_2020}), allow uniform rotation to set in, leading to a delayed collapse to a black hole. Other mechanisms that enforce uniform rotation in differentially rotating remnants are magnetic braking and shear viscosity. Finally, if $M<M_\text{max,rot}$, another mass comparison is in order, this time with $M_\text{max}$, which is the maximum mass of a cold, nonrotating star. After differential rotation is effectively damped, the configuration settles down to rigid rotation. However, since the magnetic field generated during the merger will spin the star down, when the remnant's rotational kinetic energy is dissipated  the configuration will eventually collapse to a black hole, unless $M<M_\text{max}$, in which case a cold, nonrotating neutron star will be the final outcome.

After the initial period of the remnant's formation (i.e. the first few ms), a phase that is highly nonlinear, it is possible to regard the system as a quasi-stationary, slowly drifting equilibrium state with the addition of oscillations, that are close to the linear normal modes (quasi-linear combination tones also appear) and of a spiral deformation \citep{Stergioulas_etal_2011, Bauswein_Stergioulas_2015, Bauswein_etal_2016, Bauswein_Stergioulas_2019}. Ignoring some aspects of the state of the remnant (non-axisymmetric deformations, oscillations, time-dependence, thermal structure), can lead to simplified models of its structure, with varying degree of accuracy.

In this context and depending on the two initial neutron star masses and EOS describing stellar matter, the remnant can be approximated during the various phases of its lifetime, as either a hypermassive neutron star (HMNS), or a supramassive neutron star (SMNS). In connection to the three critical masses mentioned above, a uniformly rotating star is called supramassive when its rest mass exceeds the maximum rest mass $M_\text{max}$ of a nonrotating star constructed with the same EOS \citep{Cook_etal_1992}. If it is uniformly rotating, its mass is limited by the mass-shedding limit for uniform rotation (also called Kepler limit). In the case of a HMNS the star can only be  differentially rotating. The mass of a HMNS falls in the range $M_\text{max,rot} < M < M_\text{thres}$, i.e. it is below the prompt collapse threshold and above the maximum limit for rigid rotation \citep{Baumgarte_etal_2000}.
 
Numerical simulations have been the primary tool to study compact binary coalescences and have supplied a wealth of information during the last two decades. Long-lived merger remnants (i.e. with  a lifetime greater than 10ms) have been reported by e.g. \citet{Hotokezaka_etal_2011, Sekiguchi_etal_2011, Bauswein_Janka_2012, Bauswein_etal_2012, Hotokezaka_etal_2013,Bernuzzi_etal_2014,Dietrich_etal_2015,DePietri_etal_2016,Radice_etal_2018}, whereas the remnant's rotational profile was studied extensively in a variety of setups \citep{Kastaun_Galeazzi_2015, Bauswein_Stergioulas_2015, Kastaun_etal_2016, Endrizzi_etal_2016, Kastaun_etal_2017, Ciolfi_etal_2017, Hanauske_etal_2017, Endrizzi_etal_2018, Kiuchi_etal_2018, Ciolfi_etal_2019, East_etal_2019, DePietri_etal_2020}. A common finding of the aforementioned works is that the remnant's rotation profile exhibits a maximum away from the center, that shows only a modest dependence on the exact EOS employed. This is in sharp disagreement with the differential rotation law by \citet{Komatsu_etal_1989}, hereafter KEH, which is widely used in the literature in the context of differentially rotating neutron stars and is characterised by a monotonic decline of the angular velocity $\Omega$ (for an extensive exploration of different types of equilibrium models that are possible with the KEH rotation law, see \citealt{Ansorg_etal_2009, Espino_Paschalidis_2019, Espino_etal_2019}). Rotational profiles for BNS merger remnants in eccentric mergers have been reported in \citet{Paschalidis_etal_2015, East_etal_2016}.

The need of a more realistic rotation law to construct equilibrium models of differentially rotating BNS remnants was first tackled by \citet{Galeazzi_etal_2012}, who introduced a new parametrized functional form of the rotational profile that enabled the reproduction of different power laws in the outer envelope of the angular velocity distribution. A further step was taken by \citet{Uryu_etal_2016}, who suggested a simple generalized differential rotation law, which could cover various cases using two parameters. Building up on this work, a subsequent study by \citet{Uryu_etal_2017} made a more systematic suggestion of new formulations for modelling differential rotation of compact stars, among which were a 3-parameter and a 4-parameter rotation law. \citeauthor{Uryu_etal_2017} presented selected solutions of differentially rotating, stationary, and axisymmetric compact stars in equilibrium, as examples of their proposed rotation laws. Other works that explored departure from the common \citet{Komatsu_etal_1989} rotation law were \citet{Bauswein_Stergioulas_2017, Bozzola_etal_2018}, where a 3-parameter piecewise extension of the \citeauthor{Komatsu_etal_1989} law was used to distinguish two different regions inside the star. In this way, the envelope was allowed to rotate more slowly, so that stars could reach higher masses (typical for remnants) without encountering mass-shedding.

For the case of disks around black holes, recently, a new 4-parameter family of rotation laws has also been considered, describing single rotating toroids and toroids rotating around black holes \citep{Mach_Malec_2015}. \citet{Karkowski_etal_2018} have studied an extension of the \citeauthor{Mach_Malec_2015} law, proposing a general relativistic Keplerian rotation law for self-gravitating disks around spinning black holes. Furthermore, a generalisation of the above laws to non-Keplerian rotation has been suggested by \citet{Kulczycki_Malec_2020}.

After the work of \citet{Uryu_etal_2017}, several authors started investigating the newly suggested laws for modelling realistic rotational profiles of differentially rotating, relativistic stars. \citet{Zhou_etal_2019} employed the 4-parameter nonmonotonic rotation law of \citeauthor{Uryu_etal_2017} to construct differentially rotating strange star models investigating their maximum mass, toroidal solutions and the
relationship between the threshold mass for prompt collapse to a black hole and angular momentum. \citet{Passamonti_Andersson_2020} studied the 3-parameter rotation law of \citeauthor{Uryu_etal_2017} and its impact on the low $T/|W|$ instability \citep{Watts_etal_2005}, using time evolutions of the linearised dynamical equations in Newtonian gravity, to study nonaxisymmetric oscillations and identify the unstable modes. \citet{Xie_etal_2020} continued the study of \citeauthor{Passamonti_Andersson_2020} with a fully relativistic initial survey of the nonlinear effects associated with unstable modes for different rotation laws, employing the 3-parameter rotation law of \citeauthor{Uryu_etal_2017} as well.

In light of recent works, the motivation for this study is to explore the new 4-parameter differential rotation by \citeauthor{Uryu_etal_2017} as a more realistic and versatile option for modelling compact BNS merger remnants. Our aim is to present a first systematic study of equilibrium models constructed with the new rotation law and investigate their properties.

The structure of the paper is as follows: in Section~\ref{sec:method} we present the general framework of this study. The basic equations are laid out in Section~\ref{sec:basic_eqns}, their modifications for the differential rotation law under examination are described in Section~\ref{sec:modified_eqns} and a convergence study is presented in Section~\ref{sec:convergence}. We define the equilibrium sequences that we construct in Section~\ref{sec:definitions}, compare to the corresponding KEH sequences in Section~\ref{sec:comparisons}. In Section~\ref{sec:results} we focus on the effect of different values for the rotation law's parameters, while in Section~\ref{sec:solutions_classification} we identify different types of solutions corresponding to the different rotation law parameters' values employed. Finally, in Section~\ref{sec:conclusions} we present a summary of our main results and some conclusions.

Throughout the text we employ dimensionless units for all physical quantities by setting $c = G = M_\odot = 1 $.

\section{Method}
\label{sec:method}
\subsection{Framework and basic equations}
\label{sec:basic_eqns}
For this work, we construct stationary, axisymmetric equilibrium stellar models, with matter described as a perfect fluid.  We refer to \citet{Friedman_Stergioulas_2013, Paschalidis_Stergioulas_2017}, for a more detailed exposition.

The line element for a stationary, axisymmetric star in full general relativity (GR) can be written in the form
\begin{equation}
ds^2 = -e^{\gamma + \rho} dt^2 + e^{\gamma - \rho} r^2 \sin^2 
\theta (d\phi - \omega dt)^2 + e^{2\mu} (dr^2 + r^2 d\theta^2) \label{eq:stationary_axisym_metric} \, , 
\end{equation}
where $ \gamma $, $ \rho $, $ \omega $ and $ \mu $ are metric functions depending only on the coordinates $ r $ and $ \theta $.
We assume the usual polytropic EOS 
\begin{equation}
p = K \rho ^{1 + \frac{1}{N}} \label{eq:polytropic_EOS} \, ,
\end{equation}
for which the first law of thermodynamics gives
\begin{equation}
\epsilon = \rho + N p \label{eq:energy_density} \, ,
\end{equation}
where $p$ is the pressure, $ \rho $ is the rest mass density, $\epsilon$ is the energy density, $ K $ is the polytropic constant and $ N $ is the polytropic index. Note that one should not confuse the metric function $\rho$ with the rest mass density, since the quantity meant by the symbol "$\rho$" should be evident from the relative context. 

In order to create our equilibrium models we use the {\tt rns} code \citep{Stergioulas_Friedman_1995}. The code is based on the KEH scheme \citep{Komatsu_etal_1989} and includes modifications suggested by \citet{Cook_etal_1992}. 

In all KEH-type codes, starting from an initial guess for the metric potentials $ \gamma $, $ \rho $, $ \omega $, $ \mu $, the energy density $\epsilon$ and the angular velocity $\Omega$, one uses the first integral of the hydrostationary equilibrium and the specified EOS to obtain an updated matter distribution. Subsequently, the $ \gamma $, $ \rho $, $ \omega $ and $ \mu $ distributions are updated and the steps are resumed until convergence to a solution is achieved. One of the basic steps of the scheme is to find a new angular velocity distribution in each iteration. Initially, we need a new value for the angular velocity at the equator, $\Omega_e$, then for the angular velocity at the rotation axis, $\Omega_c$ and finally a new $\Omega$ distribution everywhere inside the star. 

In order to tackle these requirements, we need to start with the equation of motion. The projection of the conservation  of the stress-energy tensor orthogonal to the 4-velocity $u^{\alpha}$ is
\begin{equation}
q^{\alpha}_{\;\; \gamma} \nabla_{\, \beta} T^{\, \beta \gamma} = 0 \label{eq:proj_conserv_str_enrg_tens} \; , 
\end{equation}
where
\begin{equation}
q^{\alpha \beta} = g^{\alpha \beta} + u^{\alpha} u^{\beta} \label{eq:projection_operator}
\end{equation}
is the projection operator orthogonal to $u^{\alpha}$, with $g_{\alpha \beta}$ being the spacetime metric and 
\begin{equation}
T^{\alpha \beta} = \epsilon u^{\alpha} u^{\beta} + p q^{\alpha \beta}  \label{eq:stress_enrg_tensor}
\end{equation}
the stress-energy tensor. The projection \eqref{eq:proj_conserv_str_enrg_tens} gives rise to the relativistic Euler equation, which for the case of a stationary, axisymmetric star has the form
\begin{align}
\frac{\nabla_{\alpha} p}{(\epsilon + p)} & = - u^{\beta} \nabla_{\beta} u_{\alpha} \nonumber  \\ 
&= \nabla_{\alpha} \ln u^t - u^t u_{\phi} \nabla_{\alpha} \Omega \label{eq:relat_euler_eq} \; . 
\end{align}
For barotropes, since $\epsilon = \epsilon(p)$, we can define a function
\begin{equation}
H(p) := \int_0^p \frac{dp'}{\epsilon(p')+p'} \label{eq:H_function} \; , 
\end{equation}
that satisfies
\begin{equation}
\nabla H = \nabla \ln h - \frac{T}{h} \nabla s \label{eq:H_satisfies} \; , 
\end{equation}
where $h = (\epsilon + p) / \rho$ is the specific enthalphy (i.e. the enthalpy per unit rest mass), $T$ is the temperature and $s$ is the specific entropy (i.e. the entropy per unit rest mass). Taking into account \eqref{eq:H_function}, the Euler equation \eqref{eq:relat_euler_eq} is written as
\begin{equation}
\nabla (H - \ln u^t) = - F \nabla \Omega \label{eq:hydroequil_v1} \; , 
\end{equation}
where $F = u^t u_{\phi} $, i.e. $F$ denotes the gravitationally redshifted angular momentum per unit rest mass and enthalpy. If we consider stars with a homogeneous entropy distribution, then from \eqref{eq:H_satisfies} it follows that, up to a constant, $H = \ln h$ and the hydrostationary equilibrium equation is written as
\begin{equation}
\nabla \left( \ln \frac{h}{u^t}\right) = - F \nabla \Omega \label{eq:hydroequil_v2} \; . 
\end{equation}
In the case of differential rotation, $F = F(\Omega)$ and Equation \eqref{eq:hydroequil_v1} gives
\begin{equation}
 H - \ln u^t + \int_{\Omega_\text{pole}}^{\Omega}F(\Omega') d\Omega' = \text{constant} \label{eq:FIHE} \; , 
\end{equation}
where the lower $\Omega$ integration limit is chosen as the value of $\Omega$ at the pole, where $H$ and the 3-velocity $\upsilon$ measured in the frame of a zero angular momentum observer (ZAMO) vanish.

In order to obtain a new value for the angular velocity at the equator $\Omega_e$, we need to equate the first integral of the hydrostationary equilibrium \eqref{eq:FIHE} at the equator to its value at the pole and solve the resulting expression numerically. This step involves making a choice for the differential rotation law describing the models that we construct, namely the function $F(\Omega)$.

Until recently, the majority of works that tackled equilibrium modelling and differential rotation followed KEH in adopting a simple, computationally convenient, differential rotation law of the form,  
\begin{equation}
F(\Omega) = A^2 (\Omega_c - \Omega),  \label{eq:KEH_rotlaw}
\end{equation}
where $ A $ is a positive constant that determines the length scale over which the angular velocity $\Omega$ changes within the star and $ \Omega_c $ is the angular velocity at the center of the configuration. Rotation law \eqref{eq:KEH_rotlaw} is reduced to uniform rotation for $A \rightarrow \infty$ and to the $j$-constant law for $A \rightarrow 0$. Here, $j= h u_\phi$ is the specific angular momentum, i.e. the angular momentum per unit baryon mass, consistent with the integral expression of the total angular momentum $J = \int j dM_0$. We note that along an axisymmetric flow the specific angular momentum $j$ is conserved, i.e. its Lie derivative vanishes, $\mathcal{L}_u\left(h u_\phi\right) = 0$.

For this study, the version of the {\tt rns} code used in \citet{Stergioulas_etal_2004} was expanded in order to implement the 4-parameter rotation law introduced in \citet{Uryu_etal_2017}
\begin{equation}
\Omega = \Omega_c \frac{1 + \left( \dfrac{F}{B^2 \Omega_c} \right)^p}{1 + \left( \dfrac{F}{A^2 \Omega_c} \right)^{q+p}} \, , \label{eq:Uryuetal_rotlaw8}
\end{equation}
(hereafter Uryu+ law), where $p$ controls the growth of the rotation curve near the rotation axis and $q$ controls the asymptotic behavior of $\Omega(r)$  (setting $q = 3$ results in recovering the Keplerian rotation law in the Newtonian limit). In this work, we choose the values $\{p, q\} = \{1, 3\}$, for which the integral in \eqref{eq:FIHE} can be calculated analytically.

Here, we should point out that in the case of the transient compact remnant formed after a BNS merger, the numerically extracted $\Omega(r)$ profile in the equatorial plane, implies that $F(\Omega)$ is not a one-to-one function,   In order to overcome this problem, we need to consider  $\Omega = \Omega(F)$, instead of $F= F(\Omega)$ \citep{Uryu_etal_2017}. This means that the integral in Equation~\eqref{eq:FIHE} must be rearranged as
\begin{equation}
\int F d\Omega = \int F \frac{d\Omega}{dF} dF  \; . \label{eq:integrability_cond}
\end{equation}

As far as the parameters $A$, $B$ in the rotation law \eqref{eq:Uryuetal_rotlaw8} are concerned, we follow \citet{Uryu_etal_2017, Zhou_etal_2019} and instead of explicitly exploring different values for A and B, we opt to fix the ratios of the maximum angular velocity over the angular velocity at the center of the configuration, $\Omega_\text{max} / \Omega_c$, and of the angular velocity at the equator over the angular velocity at the center, $\Omega_e / \Omega_c$, to certain selected values. We determine the value of the parameters $A$ and $B$ by solving for them in each iteration. Furthermore, we implement the aforementioned $\Omega$ ratios as extra input parameters in the {\tt rns} code, so that we can investigate their influence on representative equilibrium models of each sequence (\autoref{tab:lambda12_study}). We denote these parameters as
\begin{equation}
\lambda_1 =  \frac{\Omega_\text{max}}{\Omega_c}  \label{eq:lambda1},
\end{equation}
\begin{equation}
\lambda_2 = \frac{\Omega_e}{\Omega_c}   \label{eq:lambda2}.
\end{equation}
We note that as reference values for the ratios $\{\lambda_1, \lambda_2 \}$ we adopt the choice $\{ 2.0, 0.5 \}$ as in \citet{Uryu_etal_2017}. Equilibrium models reported in Tables~\ref{tab:convergence_study}, \ref{tab:physical_quantities} and \ref{tab:CFC_study} are computed using these reference values, whereas later we also explore other values of $\{\lambda_1, \lambda_2 \}$ both for selected equilibrium models (Table~\ref{tab:lambda12_study}) and full sequences (Table~\ref{tab:seqC_variants}).

\subsection{Implementation of the Uryu+ rotation law}
\label{sec:modified_eqns}
Having specified the differential rotation law we are interested in, we can return to the first integral of the hydrostationary equilibrium \eqref{eq:FIHE} and apply it at the pole and equator, in order to compute a new value for $\Omega_e$. We substitute the following expressions for the 4-velocity component $u^t$ and the 3-velocity $\upsilon$
\begin{equation}
u^t = \dfrac{e^{-(\gamma + \rho)/2}}{\sqrt{1-\upsilon^2}}, \label{eq:ut}
\end{equation}
and
\begin{equation}
\upsilon = (\Omega - \omega) r \sin \theta e^{-\rho}, \label{eq:3velocity}
\end{equation}
in \eqref{eq:FIHE} to obtain:
\begin{align}
\frac{1}{2} \left( \gamma_e + \rho_e -\gamma_p - \rho_p \right) &+ \frac{1}{2} \ln \left[ 1- \left( \Omega_e - \omega_e \right)^2 r_e^2 e^{-2\rho_e}  \right] \nonumber \\ 
&+ \int_{0}^{F_e} F \frac{d\Omega}{dF}   d F = 0. \label{eq:hydro_equil_int_pole_equator} 
\end{align}
In the above equation, we need to differentiate the $\Omega(F)$ law, calculating the law's parameters in the process. We solve the system of equations for ratios $\lambda_1$ and $\lambda_2$ to determine expressions for parameters A and B
\begin{equation}
\lambda_1 = \frac{1 + \left( \dfrac{F_\text{max}}{B^2 \Omega_c} \right)^p}{1 + \left( \dfrac{F_\text{max}}{A^2 \Omega_c} \right)^{q+p} }, \label{eq:ABsystem_eqlambda1gen}
\end{equation}
\begin{equation}
\lambda_2 = \frac{1 + \left( \dfrac{F_e}{B^2 \Omega_c} \right)^p}{1 + \left( \dfrac{F_e}{A^2 \Omega_c} \right)^{q+p}} \, ,  \label{eq:ABsystem_eqlambda2gen}
\end{equation}
and for the general $\{p, q \}$ case we find
\begin{equation}
A^2 = \left[ \frac{\left( F_e F_\text{max} \right)^p}{{\Omega_c}^{p+q}}  \frac{\left( \lambda_2 \: {F_e} ^q - \lambda_1 \: {F_\text{max}}^q \right)}{\left[ \: (\lambda_1 - 1) \: {F_e}^p - (\lambda_2 -1) \: {F_\text{max}}^p \: \right]} \right]^{  \frac{1}{p+q} } \, , \label{eq:Asq_general}
\end{equation}
and
\begin{equation}
B^2 = \frac{F_e F_\text{max}}{\Omega_c} \left[ \frac{\left( \lambda_2 \: {F_e} ^q - \lambda_1 \: {F_\text{max}}^q \right)}{ \left[ \: \lambda_2 \: (\lambda_1 -1) \: {F_e}^{p+q} - \lambda_1 \: (\lambda_2 - 1) \: {F_\text{max}}^{p+q} \: \right] } \right]^{1/p}  . \label{eq:Bsq_general}
\end{equation}
Assuming that we need the above expressions to be defined in the real numbers domain, a relevant constraint must be fulfilled, due to the numerator of both fractions in \eqref{eq:Asq_general} and \eqref{eq:Bsq_general}, namely
\begin{equation}
F_e > \left( \frac{\lambda_1}{\lambda_2} \right)^{1/q} F_\text{max}  \; . \label{eq:Fe_Fmax_constraint_lambdagen}
\end{equation}
Having secured analytic expressions for the parameters A and B, we differentiate $\Omega(F)$ from expression \eqref{eq:Uryuetal_rotlaw8}, calculate the integral \eqref{eq:integrability_cond}, substitute into \eqref{eq:hydro_equil_int_pole_equator} and use logarithmic and trigonometric identities to get the expression (for the case that $\{p,q\}=\{1,3\}$)
\begin{align}
& \left( \gamma_e + \rho_e -\gamma_p - \rho_p \right) + \ln \left[ 1- \left( \Omega_e - \omega_e \right)^2  r_e^2 e^{-2\rho_e}  \right] = -2 F_e \Omega_e \nonumber \\ 
&+ \frac{A^2 {\Omega_c}^2}{2} \left\lbrace 2 \frac{A^2}{B^2} \arctan \left( \dfrac{{F_e}^2 }{ A^4 {\Omega_c}^2} \right)  - \sqrt{2} \left[ \arctan\left( 1 - \frac{F_e \sqrt{2}}{A^2 \Omega_c} \right) \right. \right. \nonumber \\
& \left. \left. -  \arctan \left( 1 + \frac{F_e \sqrt{2}}{A^2 \Omega_c} \right) \right]
 + \sqrt{2}\tanh^{-1} \left( \frac{ A^2 \Omega_c F_e \sqrt{2}}{{F_e}^2 + A^4 {\Omega_c}^2} \right) \right\rbrace \; . \label{eq:hydroequil_numsolve}
\end{align}
The above equation needs to be solved numerically, in order to obtain an updated value for $\Omega_e$. Equation~\eqref{eq:hydroequil_numsolve} involves occurrences of $F_e$ and $\Omega_c$, which need to be expressed in terms of $\Omega_e$, so that a numerical solution can properly take place. The former can be eliminated using the expression $F = u^t u_{\phi}$,
\begin{equation}
F = \frac{\upsilon^2}{\left(1 -\upsilon^2 \right) \left( \Omega - \omega \right)}= \frac{\left( \Omega - \omega \right) r^2 \sin^2 \theta e^{-2\rho}}{\left[ 1 - \left( \Omega - \omega \right)^2 r^2 \sin^2 \theta e^{-2\rho} \right]} \, , \label{eq:F_expr_general}
\end{equation}
applied at the equator
\begin{equation}
F_e = \frac{\left( \Omega_e - \omega_e \right) {r_e}^2 e^{-2\rho_e}}{\left[ 1 - \left( \Omega_e - \omega_e \right)^2 {r_e}^2 e^{-2\rho_e} \right]} \; .
 \label{eq:kill_Fe}
\end{equation}
The latter can be replaced using the definition of $\lambda_2$  \eqref{eq:lambda2}
\begin{equation}
\Omega_c = \frac{1}{\lambda_2} \; \Omega_e \; , \label{eq:kill_Omegac}
\end{equation}
so that in the end, on both hand sides of \eqref{eq:hydroequil_numsolve} only the quantity $\Omega_e$ is encountered.

For the sake of completeness, we should point out that after obtaining a new value for $\Omega_e$ through the solution of \eqref{eq:hydroequil_numsolve}, a new value for $\Omega_c$ can be obtained through \eqref{eq:kill_Omegac} and a new distribution of $\Omega$ is obtained everywhere inside the star by numerically solving the expression
\begin{equation}
\Omega = \Omega_c \, \frac{1 + \left( \dfrac{\left( \Omega - \omega \right) r^2 \sin^2 \theta e^{-2\rho}}{ B^2 \Omega_c \left[ 1 - \left( \Omega - \omega \right)^2 r^2 \sin^2 \theta e^{-2\rho} \right]} \right)^p}{1 + \left( \dfrac{\left( \Omega - \omega \right) r^2 \sin^2 \theta e^{-2\rho}}{A^2 \Omega_c \left[ 1 - \left( \Omega - \omega \right)^2 r^2 \sin^2 \theta e^{-2\rho} \right]} \right)^{p+q}} \; ,  \label{eq:Omega_everywhere_pq}
\end{equation}
which emerges by combining equations~\eqref{eq:Uryuetal_rotlaw8} and \eqref{eq:F_expr_general}.

With an updated $\Omega$ distribution at hand, courtesy of \eqref{eq:Omega_everywhere_pq}, the iterative algorithm moves on to update the matter distribution, a step for which we need a new enthalpy relation compatible with the new Uryu+ rotation law \eqref{eq:Uryuetal_rotlaw8}. In order to accomplish this, we solve the first integral of the hydrostationary equilibrium and equate its value at an arbitrary point anywhere inside the star to its value at the pole. In the context of the fixed-point iteration in the KEH scheme, the following expression applies generally, regardless of which specific rotation law we choose
\begin{align}
H = H_\text{surface} + \ln \frac{u^t}{u_\text{pole}^t} - \int_{\Omega_{c}}^{\Omega} F(\Omega') d\Omega' \; . \label{eq:new_H_distr}
\end{align}
For the case of the Uryu+ law in particular, the integral in \eqref{eq:new_H_distr} is given by the expression (we note that its calculation was already required to obtain Equation~\eqref{eq:hydroequil_numsolve})
\begin{align}
\int_{\Omega_{c}}^{\Omega} F(\Omega') d\Omega' & =  \int_{0}^{F} F' \frac{d\Omega}{dF'} dF'= \nonumber \\
& F \Omega - \frac{A^2 {\Omega_c}^2}{4} \left\lbrace \frac{2 A^2}{B^2} \arctan \left( \dfrac{F^2 }{ A^4 {\Omega_c}^2} \right)  \right. \nonumber \\
& - \sqrt{2} \left[ \arctan\left( 1 - \frac{F \sqrt{2}}{A^2 \Omega_c} \right) -  \arctan \left( 1 + \frac{F \sqrt{2}}{A^2 \Omega_c} \right) \right] \nonumber \\
& + \left. \sqrt{2}\tanh^{-1} \left( \frac{ A^2 \Omega_c F \sqrt{2}}{{F}^2 + A^4 {\Omega_c}^2} \right) \right\rbrace \; , \label{eq:Uryu_law_integral}
\end{align} 
where \eqref{eq:F_expr_general} was used.

\begingroup
\setlength{\tabcolsep}{4.7pt}
\begin{table*}
        \centering
        \caption{Convergence study for a rapidly rotating model of high compactness, constructed with the Uryu+ rotation law \eqref{eq:Uryuetal_rotlaw8}. The model has a central energy density of $\epsilon_c = 3.3 \times 10^{-3}$ and an axis ratio of $r_p/r_e = 0.43$ (this is model C6 calculated with $\{\lambda_1, \lambda_2\} = \{2.0,0.5\}$ in Section~\ref{sec:equilibriums} and Table~\ref{tab:seqC_variants}). The first column shows the resolution used in the $(s,\mu)$ grid of the {\tt rns} code. The remaining columns show the maximum energy density $\epsilon_\text{max}$, the gravitational mass $M$, the rest mass $M_0$, the angular momentum $J$, the ratio of the rotational kinetic energy $T$ over the absolute value of the gravitational binding energy $|W|$, the angular velocity at the rotation axis $\Omega_c$, the maximum value of angular velocity $\Omega_\text{max}$, the angular velocity at the equator $\Omega_e$, the angular velocity of a free particle in circular orbit at the equator $\Omega_K$, the circumferential radius $R_e$, the coordinate radius $r_e$ at the equator and the 3-dimensional general relativistic virial index GRV3. All quantities are reported in dimensionless units $c = G = M_\odot = 1 $, in which we choose $K=100$ for the $N=1$ polytropic EOS. The last line displays the convergence order, based on the last three resolutions, except for $M, M_0$, for which the first three resolutions were used.}
        \label{tab:convergence_study}
        \begin{tabular}{lcccccccccccc}
                \hline
                resolution & $\epsilon_\text{max}$ & $M$ & $M_0$ & J & $T/|W|$ & $\Omega_c$ & $\Omega_\text{max}$ & $\Omega_e$ & $\Omega_K$ & $R_e$ & $r_e$ & GRV3\\
                {\tt SDIV} $\times$ {\tt MDIV} & $\left(\times 10^{-3}\right)$ & & & & $\left(\times 10^{-1}\right)$ & $\left(\times 10^{-2}\right)$ & $\left(\times 10^{-2}\right)$ & $\left(\times 10^{-2}\right)$ & $\left(\times 10^{-2}\right)$ & & & $\left(\times 10^{-5}\right)$\\
                \hline
                $201 \times 101$ & 3.78409 & 2.27105755 & 2.477858 & 3.95626 & 1.93275 & 7.22571 & 14.4641 & 3.61286 & 5.98030 & 8.47078 & 5.80855 & 24.3253\\
		        $401 \times 201$ & 3.78249 & 2.27111885 & 2.477878 & 3.95707 & 1.93325 & 7.22415 & 14.4497 & 3.61208 & 5.97857 & 8.47236 & 5.81006 & 8.84179\\
	        	$801 \times 401$ & 3.78169 & 2.27113957 & 2.477881 & 3.95736 & 1.93348 & 7.22363 & 14.4473 & 3.61181 & 5.97797 & 8.47292 & 5.81060 & 5.31201\\
	        	$1601 \times 801$ & 3.78156 & 2.27113961 & 2.477876 & 3.95742 & 1.93353 & 7.22351 & 14.4470 & 3.61175 & 5.97783 & 8.47305 & 5.81073 & 4.15341\\		
		        \hline
		        conv. order & 2.5 & 1.7 & 2.9 & 2.2 & 2.1 & 2.1 & 2.8 & 2.1 & 2.1 & 2.1 & 2.0\\
		        \hline
        \end{tabular}
\end{table*}
\endgroup

Here, we should point out that while running the extended {\tt rns} code, we chose to keep the central energy density fixed instead of the maximum energy density. This was an experimental choice that after several trials was found to improve model convergence. However, if we want to be thorough in our analysis, we need to obtain a new relation for calculating $r_e$ for the case that we opt to keep the maximum energy density fixed. The derivation is based on calculating the first integral of the hydrostationary equilibrium \eqref{eq:FIHE} at the location of the maximum energy density, equating its value there, to its value at the pole and solving for $r_e$. The analysis is general and in the current framework it holds regardless of which rotation law we choose (in a similar way as in the derivation of the enthalpy relation). The expression we obtain is  
\begin{equation}
{r_e}^2 = \frac{2 H_\text{max} + \ln \left( 1 - {\upsilon^2_\text{max}} \right) + 2 \int_{\Omega_{c}}^{\Omega_\text{max}} F(\Omega) d\Omega }{\hat{\gamma}_p + \hat{\rho}_p - \hat{\gamma}_\text{max} -\hat{\rho}_\text{max}
}, \label{eq:new_re_polemax_gen}
\end{equation}
where all occurrences of the subscript "max" in Equation~\eqref{eq:new_re_polemax_gen} denote the corresponding quantities at the location where the density has its maximum value (note that, in contrast to the rest of this paper, even $\Omega_{\rm max}$ is to be interpreted in this way in the above equation). Furthermore, in Equation~\eqref{eq:new_re_polemax_gen} quantities denoted with a "\^{}" are rescaled with $r_e$ as described in Appendix~\ref{sec:appendix_rescalings}. Replacing the integral term in Equation~\eqref{eq:new_re_polemax_gen} with the corresponding calculation for the specific rotation law under study (e.g. KEH, Uryu+, etc.) yields the relevant relation for each case.  For the case of the Uryu+ law, the integral in \eqref{eq:new_re_polemax_gen} is calculated at the location of the energy density maximum via \eqref{eq:Uryu_law_integral}.

We need to stress that a generalization of the aforementioned procedure to arbitrary $\{p,q\}$ values is not straightforward. In case we want to study a slightly different rotation law, but still described by \eqref{eq:Uryuetal_rotlaw8}, e.g. $\{p,q\}$ values other than $\{1, 3\}$, things are not so simple as changing the parameters $\{p,q\}$ to our desired values. The obvious reason is that the integral $\int F'\frac{d\Omega}{dF'}dF'$ would have to be recalculated for the new $\{p,q\}$ values and all the relevant equations where it is involved (i.e. expressions for $\Omega_e$, enthalpy and $r_e$ in the case that the maximum energy density is held fixed), would have to be modified.

As a demonstration, attempting to come up with a general expression for the integral $\int F'\frac{d\Omega}{dF'}dF'$ by leaving the differential rotation law \eqref{eq:Uryuetal_rotlaw8} in its general form (without using specific values for $\{p,q\}$), leads to occurrences of the incomplete beta function
\begin{equation}
B_z (a,b) = \int_0^z t^{a-1} (1-t)^{b-1} dt. \label{eq:incomp_beta_func}
\end{equation}
This adds to the complexity of the problem, compared to the simpler case of setting $\{p,q\} = \{1,3\}$. Therefore, a thorough investigation of the possible $\{p,q\}$ range, may prove challenging if one additionally considers non-integer $\{p,q\}$ values.

\subsection{EOS and convergence study}
\label{sec:convergence}

In the present study, we choose $N=1$ and $K=100$, which is a common choice in the literature for testing numerical codes. Note that for the simple polytropic EOS the results can be rescaled to any other choice of $K$. In practice, we construct a polytropic model with $K=1$ and then rescale the results to $K=100$ ($K^{N/2}$ has units of length).  The maximum mass nonrotating model has a central energy density $\epsilon_c = 4.122 \times 10^{-3}$, with a gravitational mass of $M \simeq 1.64$ and a rest mass of $M_0 \simeq 1.8$. For the maximum mass uniformly rotating model (i.e. at the mass-shedding limit) the corresponding values are: $\epsilon_c = 3.34 \times 10^{-3}$ with gravitational mass $M \simeq 1.88$ and rest mass $ M_0 \simeq 2.07 $. We also note that the nonrotating model C0 has a compactness $M/R\sim 0.2$, while nonrotating models A0 and B0 have a (coinciding) compactness of $\sim 0.15$. 

The {\tt rns} code uses a uniformly-spaced numerical grid with coordinates $(s,\mu)$, where $s$ is a dimensionless, compactified radial coordinate, defined through
\begin{equation}
    r=r_e\frac{s}{1-s},
\end{equation}
($s=0$ corresponds to $r=0$, $s=1/2$ corresponds to $r=r_e$ and $s=1$ corresponds to $r=+\infty$) and $\mu=\cos\theta$.


\begin{figure*}
    \includegraphics[scale=0.43]{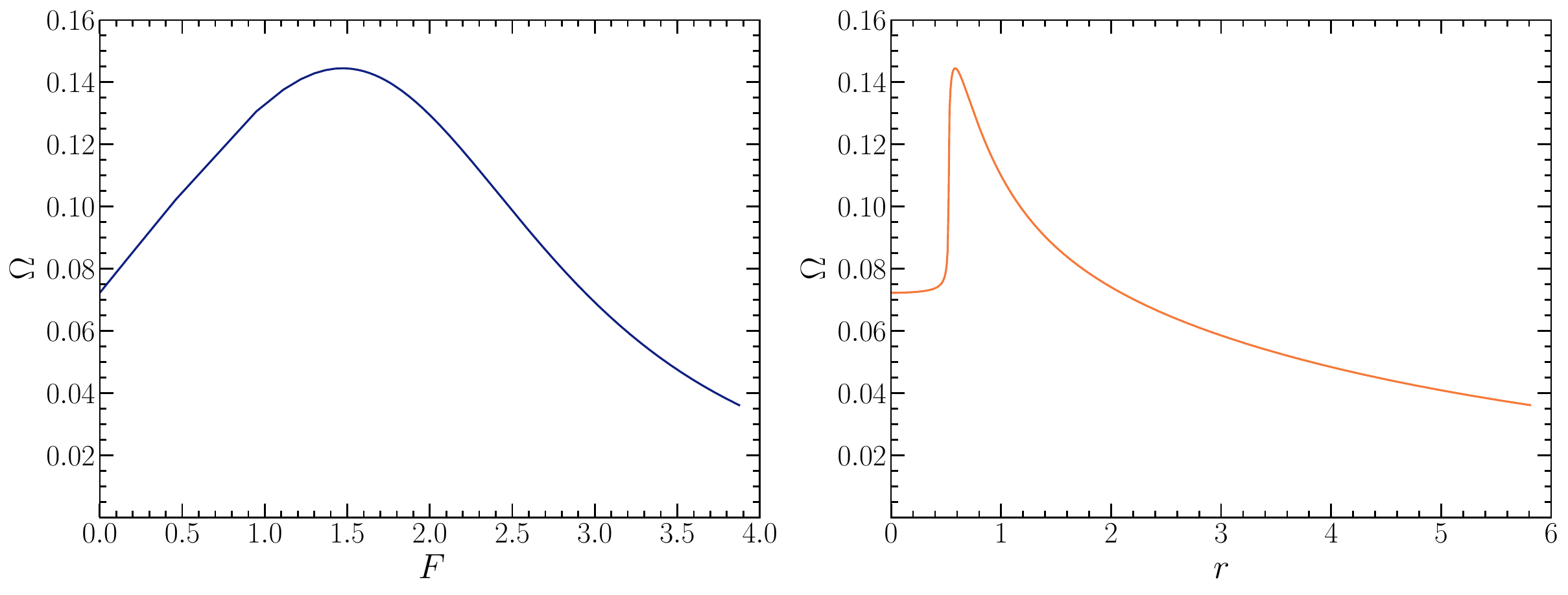}
    \caption{Angular velocity $\Omega$ profiles versus the gravitationally redshifted angular momentum per unit rest mass and enthalpy $F$ (left panel) and the coordinate radius $r$ (right panel) in the equatorial plane for model C6 (axis ratio $r_p/r_e$ = 0.43), constructed with the Uryu+ rotation law with $\{\lambda_1, \lambda_2\} = \{2.0, 0.5\}$ (see text and Table~\ref{tab:seqC_variants} for details).}
    \label{fig:C6_Omega_Fr_profiles}
\end{figure*}

Table~\ref{tab:convergence_study} shows the results of a convergence study for a rapidly rotating model of high compactness, constructed with the Uryu+ rotation law \eqref{eq:Uryuetal_rotlaw8}. The upper left panel of Figure~\ref{fig:seqC_variations_density2Dxz} shows the rest mass density distribution in the $x-z$ plane, as reference, while in Figure~\ref{fig:C6_Omega_Fr_profiles} we present the angular velocity profiles $\Omega(F)$ and $\Omega(r)$ at the equatorial plane. The model has a central energy density of $\epsilon_c = 3.3 \times 10^{-3}$ and an axis ratio of $r_p/r_e = 0.43$ (this is model C6 in Section \ref{sec:equilibriums}). We demonstrate the convergence of the following quantities with increasing resolution: the central energy density $\epsilon_c$, the maximum energy density $\epsilon_\text{max}$, the gravitational mass $M$, the rest mass $M_0$, the angular momentum $J$ and the ratio of the rotational kinetic energy $T$ over the absolute value of the gravitational binding energy $|W|$, the angular velocity at the rotation axis $\Omega_c$, the maximum value of angular velocity $\Omega_\text{max}$, the angular velocity at the equator $\Omega_e$, the angular velocity of a free particle in circular orbit at the equator $\Omega_K$, the circumferential radius $R_e$, the coordinate radius $r_e$ at the equator and the 3-dimensional general relativistic virial index GRV3. The latter has been shown to reflect the accuracy of numerical models of stationary equilibrium configurations \citep{Nozawa_etal_1998}.

For the convergence study we use $(s,\mu)$ grids of $({\tt SDIV}\times {\tt MDIV})$=$(201\times101)$, $(401\times201)$, $(801\times401)$ and $(1601\times801)$. Our results demonstrate 2nd-order convergence, as expected for the overall numerical schemes employed in {\tt rns}. Between the two highest resolutions, the relative difference is less than $\sim2.6\times10^{-5}$ for the  integrated quantities $M, M_0, J, T/|W|$ (in particular, for the masses it is less than $\sim2\times10^{-6}$)  and less than $\sim2.3\times10^{-5}$ for the local quantities $r_e, R_e, \Omega_c, \Omega_\text{{max}}, \Omega_e, \Omega_K$. 

In the remainder of this work, all numerical models will be constructed with a standard resolution of $({\tt SDIV}\times {\tt MDIV})$=$(801\times401)$.

\section{Sequences of equilibrium models}
\label{sec:equilibriums}
 
\subsection{Definition of the equilibrium sequences}
\label{sec:definitions}

For comparison with previous work \citep{Stergioulas_etal_2004, Iosif_Stergioulas_2014}, we construct three sequences of equilibrium models using the new differential rotation law \eqref{eq:Uryuetal_rotlaw8}:

\begin{itemize}
    \item Sequence A is a constant rest mass sequence with $M_0 = 1.506$.
    \item Sequence B is a constant central energy density sequence with $\epsilon_c = 1.444 \times 10^{-3}$.
    \item Sequence C is a constant central energy density sequence with $\epsilon_c = 3.3 \times 10^{-3}$.
\end{itemize}
 We note that the central energy density value of sequence C, which is the most relativistic out of the three sequences, is determined taking into account remarks in \citet{Takami_etal_2011} and \citet{Giacomazzo_etal_2011} concerning the location of the secular and dynamical instabilities. We note that the faster rotating equilibrium models of sequence C could be both secularly and dynamically unstable. For the KEH rotation law, \citet{Weih_etal_2018} showed that the dynamical instability occurs well before the turning points of constant-angular-momentum sequences\footnote{The latter are used to locate the secular axisymmetric instability in uniformly rotating stars.}. See also \citet[section 9.2]{Friedman_Stergioulas_2013} for a summary on axisymmetric stability and turning points, where the difference between indices $\Gamma$ (governing the equilibrium EOS) and $\Gamma_1$ (governing dynamical oscillations) is highlighted. This means that for realistic (i.e. hot) stars the instability limit will not be the same as in the case of cold stars. In particular, the difference between secular and dynamical instability for hot stars could be larger than the corresponding difference for cold stars. 

\interfootnotelinepenalty=10000
Tables~\ref{tab:physical_quantities} and \ref{tab:seqC_variants} list all relevant physical quantities for the three different sequences of equilibrium models.
Since we want to compare these models to the corresponding models constructed with the KEH rotation law, we re-calculated sequences A, B (defined in \citealt{Stergioulas_etal_2004}) and C (defined in \citealt{Iosif_Stergioulas_2014}) with the KEH rotation law, using our standard resolution of $({\tt SDIV}\times {\tt MDIV})$=$(801\times401)$. For the KEH sequence A, we kept the same central energy density values and a rest mass of $M_0 = 1.506$ as in \citet{Stergioulas_etal_2004}, which resulted in slightly different $r_p / r_e$ (due to the different resolution) than those reported in \citet{Stergioulas_etal_2004}. In addition, for the KEH sequence C, we calculated an additional model with $r_p / r_e = 0.43$, in order to facilitate the comparison with the Uryu+ sequence C for the case of $\{\lambda_1, \lambda_2\}=\{2.0, 0.5\}$ (Figures~\ref{fig:seqBC_mass_Uryu_vs_KEH} and \ref{fig:seqBC_mass_vs_Re_Uryu_vs_KEH}), since the latter terminates at this axis ratio\footnote{The termination of sequence C with the Uryu+ rotation law is due to a limitation of the numerical scheme in {\tt rns} - the sequence has not reached the mass-shedding limit, as is evident from the $\Omega_e$ vs. $\Omega_K$ values of model C6 displayed in Table \ref{tab:seqC_variants}.}. The detailed properties of this additional model are displayed in Table \ref{tab:extra-C-KEH-model} in Appendix \ref{seq:extra-C-KEH-model}. In the Figures~\ref{fig:mass_vs_density_classic}, \ref{fig:surface_xz}, \ref{fig:seqA_mass_Uryu_vs_KEH}, \ref{fig:seqBC_mass_Uryu_vs_KEH}, \ref{fig:seqA_mass_vs_Re_Uryu_vs_KEH} and \ref{fig:seqBC_mass_vs_Re_Uryu_vs_KEH} that follow, all models have been constructed assuming $\{\lambda_1, \lambda_2\} = \{2.0,0.5\}$.

\begingroup
\setlength{\tabcolsep}{3.4pt}
\begin{table*}
        \centering
        \caption{Physical quantities for sequences A (with constant $M_0 = 1.506$) and B (with constant  $\epsilon_c = 1.444 \times 10^{-3}$) calculated using the Uryu+ differential rotation law \eqref{eq:Uryuetal_rotlaw8}. The different quantities are defined as in Table \ref{tab:convergence_study}.}
        \label{tab:physical_quantities}
        \begin{tabular}{lcccccccccccccccc}
                \hline
                Model & $r_p/r_e$ & $\epsilon_c$ & $\epsilon_\text{max}$ & $M$ & $M_0$ & J & $T/|W|$ & $\Omega_c$ & $\Omega_\text{max}$ & $\Omega_e$ & $\Omega_K$ & $R_e$ & $r_e$ & GRV3\\
                & & $\left(\times 10^{-3}\right)$ & $\left(\times 10^{-3}\right)$ & & & & $\left(\times 10^{-1}\right)$ & $\left(\times 10^{-2}\right)$ & $\left(\times 10^{-2}\right)$ & $\left(\times 10^{-2}\right)$ & $\left(\times 10^{-2}\right)$ & & & $\left(\times 10^{-5}\right)$\\
                \hline
                A0 & 1.0 & 1.444 & 1.44400 & 1.40021 & 1.50624 & 0.0 & 0.0 & 0.0 & 0.0 & 0.0 & 3.98735 & 9.58537 & 8.12483  & 4.94068\\
                A1 & 0.9350 & 1.300 & 1.30000 & 1.40336 & 1.50600 & 0.48978 & 0.16876 & 1.13141 & 2.26283 & 0.56571 & 3.76434 & 9.94733 & 8.47659 & 7.78180\\
                A2 & 0.8834 & 1.187 & 1.18700 & 1.40618 & 1.50597 & 0.67872 & 0.31154 & 1.46576 & 2.93153 & 0.73288 & 3.60338 & 10.2712 & 8.79166 & 7.34990\\
                A3 & 0.8312 & 1.074 & 1.07400 & 1.40928 & 1.50600 & 0.84681 & 0.46416 & 1.69791 & 3.39582 & 0.84896 & 3.43843 & 10.6376 & 9.14868 & 6.99791\\
                A4 & 0.7783 & 0.961 & 0.96100 & 1.41258 & 1.50595 & 1.00873 & 0.62736 & 1.86296 & 3.72593 & 0.93148 & 3.26734 & 11.0556 & 9.55670 & 6.71677\\
                A5 & 0.7243 & 0.848 & 0.84800 & 1.41624 & 1.50599 & 1.17284 & 0.80274 & 1.97572 & 3.95145 & 0.98786 & 3.08876 & 11.5380 & 10.0283 & 6.47831\\
                A6 & 0.6689 & 0.735 & 0.73500 & 1.42027 & 1.50605 & 1.34476 & 0.99147 & 2.04215 & 4.08430 & 1.02107 & 2.90170 & 12.1014 & 10.5798 & 6.30968\\
                A7 & 0.6117 & 0.622 & 0.62200 & 1.42456 & 1.50595 & 1.52957 & 1.19443 & 2.06431 & 4.12863 & 1.03216 & 2.70530 & 12.7680 & 11.2335 & 6.16661\\
                A8 & 0.5519 & 0.509 & 0.50900 & 1.42932 & 1.50586 & 1.73385 & 1.41258 & 2.04253 & 4.08507 & 1.02127 & 2.49909 & 13.5694 & 12.0206 & 6.05597\\
                A9 & 0.4882 & 0.396 & 0.40086 & 1.43493 & 1.50612 & 1.96591 & 1.64514 & 1.97549 & 3.95098 & 0.98775 & 2.28329 & 14.5499 & 12.9843 & 6.03066\\
                A10 & 0.4183 & 0.283 & 0.32884 & 1.44091 & 1.50615 & 2.23263 & 1.88559 & 1.86009 & 3.72018 & 0.93005 & 2.05902 & 15.7684 & 14.1836 & 6.17384\\
                A11 & 0.3356 & 0.170 & 0.26606 & 1.44762 & 1.50639 & 2.54338 & 2.11646 & 1.69452 & 3.38903 & 0.84726 & 1.82984 & 17.3064 & 15.6984 & 6.57445\\
                A12 & 0.2798 & 0.110 & 0.23676 & 1.45080 & 1.50595 & 2.72584 & 2.22139 & 1.58573 & 3.17146 & 0.79287 & 1.70767 & 18.2887 & 16.6668 & 6.72845\\ \\
                B0 & 1.0 & 1.444 & 1.44400 & 1.40021 & 1.50624 & 0.0 & 0.0 & 0.0 & 0.0 & 0.0 & 3.98735 & 9.58537 & 8.12483 & 4.94068\\
                B1 & 0.950 & 1.444 & 1.44400 & 1.43873 & 1.54893 & 0.44293 & 0.12874 & 1.04200 & 2.08400 & 0.52100 & 3.94463 & 9.71434 & 8.20529 & 7.95499\\
                B2 & 0.900 & 1.444 & 1.44400 & 1.48252 & 1.59755 & 0.67619 & 0.26495 & 1.49286 & 2.98572 & 0.74643 & 3.92705 & 9.85083 & 8.28676 & 7.58352\\
                B3 & 0.849 & 1.444 & 1.44400 & 1.53373 & 1.65451 & 0.90387 & 0.41202 & 1.86097 & 3.72193 & 0.93048 & 3.91534 & 9.99772 & 8.36948 & 7.21883\\
                B4 & 0.800 & 1.444 & 1.44400 & 1.59038 & 1.71769 & 1.13525 & 0.56153 & 2.17462 & 4.34924 & 1.08731 & 3.90946 & 10.1454 & 8.44627 & 6.90944\\
                B5 & 0.750 & 1.444 & 1.44400 & 1.65729 & 1.79252 & 1.39794 & 0.72279 & 2.47431 & 4.94862 & 1.23716 & 3.90997 & 10.3013 & 8.51829 & 6.63163\\
                B6 & 0.700 & 1.444 & 1.44400 & 1.73532 & 1.88009 & 1.70054 & 0.89312 & 2.76602 & 5.53204 & 1.38301 & 3.91929 & 10.4592 & 8.57816 & 6.38871\\
                B7 & 0.650 & 1.444 & 1.44400 & 1.82678 & 1.98317 & 2.05786 & 1.07256 & 3.06059 & 6.12118 & 1.53030 & 3.94107 & 10.6134 & 8.61690 & 6.18202\\
                B8 & 0.600 & 1.444 & 1.44400 & 1.93428 & 2.10493 & 2.48740 & 1.26069 & 3.37014 & 6.74029 & 1.68507 & 3.98100 & 10.7539 & 8.62050 & 5.99032\\
                B9 & 0.550 & 1.444 & 1.44400 & 2.06020 & 2.24840 & 3.00859 & 1.45629 & 3.71100 & 7.42200 & 1.85550 & 4.04785 & 10.8636 & 8.56761 & 5.81028\\
                B10 & 0.500 & 1.444 & 1.48092 & 2.20535 & 2.41490 & 3.63795 & 1.65695 & 4.10748 & 8.21495 & 2.05374 & 4.15491 & 10.9145 & 8.42730 & 5.64007\\
                B11 & 0.450 & 1.444 & 1.62233 & 2.36580 & 2.60021 & 4.37467 & 1.85864 & 4.59661 & 9.19322 & 2.29830 & 4.32118 & 10.8656 & 8.16003 & 5.47790\\
                B12 & 0.400 & 1.444 & 1.85018 & 2.52779 & 2.78786 & 5.17196 & 2.05616 & 5.23110 & 10.4622 & 2.61555 & 4.57013 & 10.6676 & 7.72955 & 5.34501\\  
                B13 & 0.340 & 1.444 & 2.29855 & 2.68602 & 2.96781 & 6.03569 & 2.28203 & 6.27108 & 12.5422 & 3.13554 & 5.00597 & 10.1890 & 6.99623 & 5.25323\\ 
                \hline
        \end{tabular}
\end{table*}
\endgroup

\begin{figure}
        \includegraphics[width=\columnwidth]{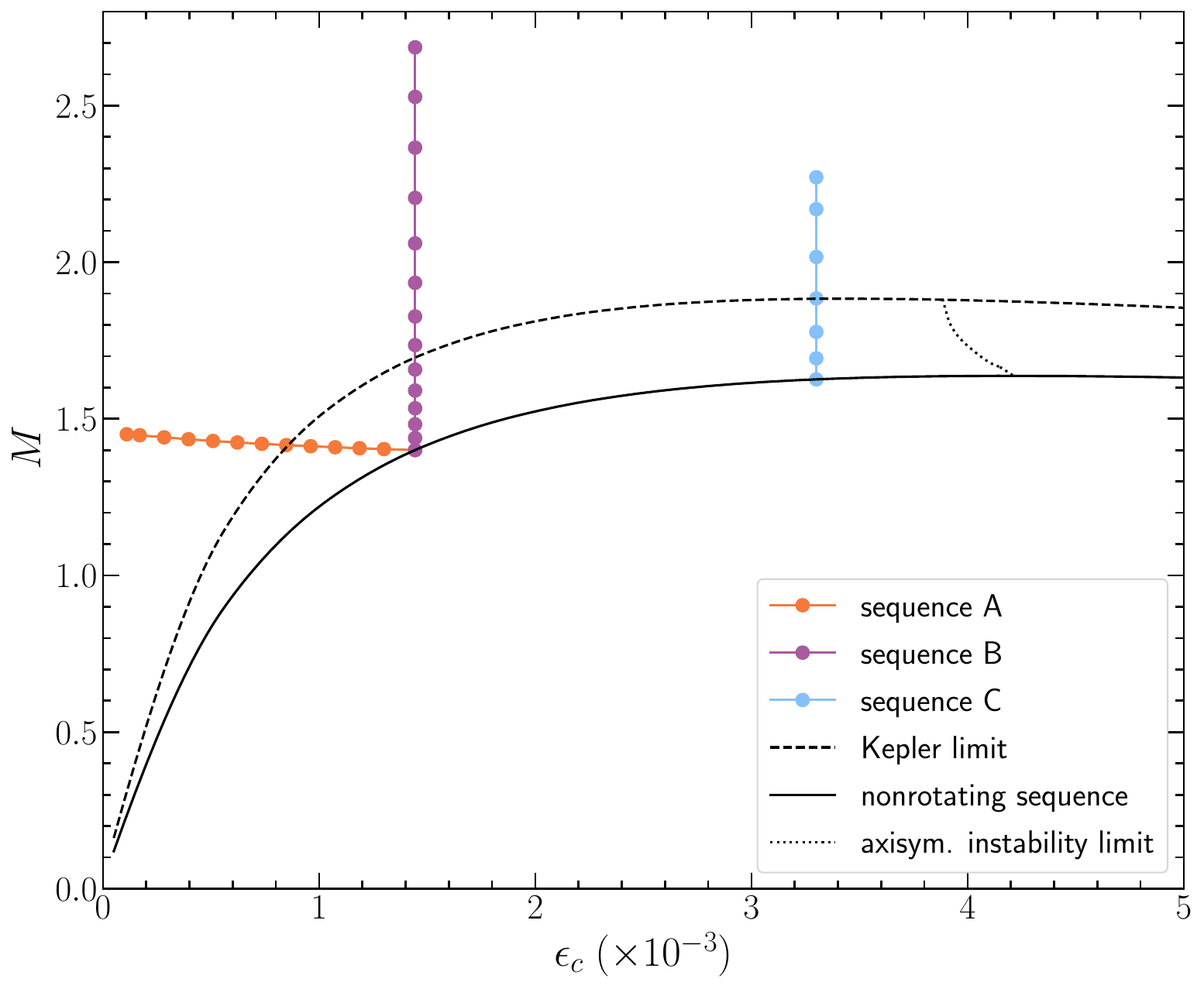}
    \caption{Gravitational mass $M$ vs. central energy density $\epsilon_c$ for sequences A, B and C. In addition, the nonrotating (TOV) sequence (solid line), the mass-shedding (Kepler) limit for uniform rotation (dashed line) and the axisymmetric instability limit for uniform rotation (dotted line) are shown.}
    \label{fig:mass_vs_density_classic}
\end{figure}

In Figure~\ref{fig:mass_vs_density_classic} we plot the gravitational mass $M$ as a function of the central energy density $\epsilon_c$ for the three sequences A, B and C. For reference, we show the TOV (nonrotating) sequence as a solid line and the mass-shedding limit for uniform rotation (Kepler limit) as a dashed line. The dotted line represents the axisymmetric stability limit for uniformly rotating models, based on the turning point method by \citet{Friedman_etal_1988}, see also models S1-S4 in \citep[Table I]{Baiotti_etal_2005}.

\subsection{Comparison to corresponding KEH sequences}
\label{sec:comparisons}

\begin{figure}
        \includegraphics[width=\columnwidth]{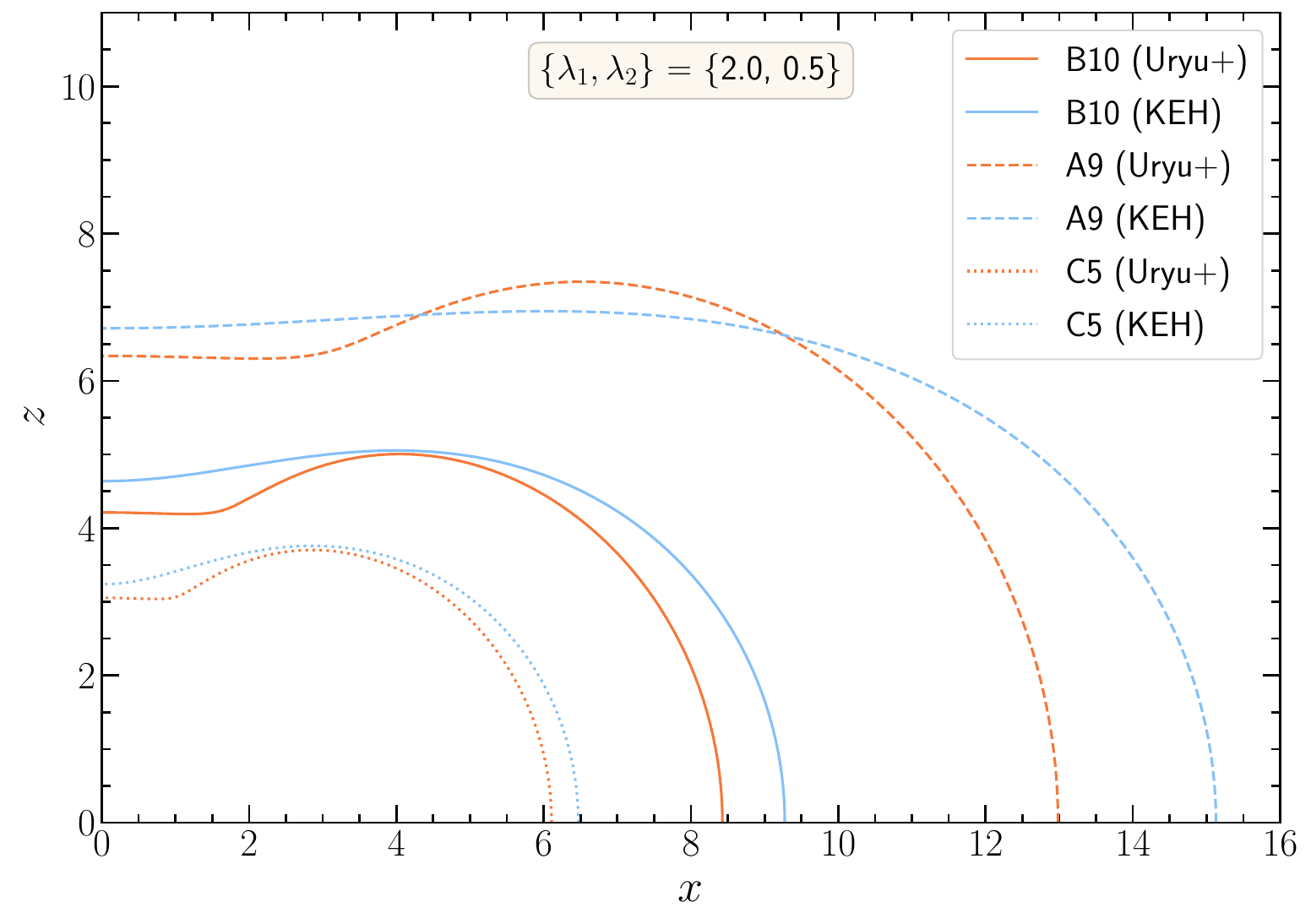}
    \caption{Stellar surfaces comparison between the Uryu+ and the KEH differential rotation laws, for representative models of each sequence with ratio $r_p / r_e \sim 0.5$. Model A9 has the same central energy density value $\epsilon_c$ for both rotation law runs, but different $r_p/r_e$ values (0.5133 for the KEH law and 0.4882 for the Uryu+ law). This is necessary, so that the constant rest mass requirement of sequence A is satisfied accurately for both rotation laws.}
    \label{fig:surface_xz}
\end{figure}

\begin{figure}
        \includegraphics[width=\columnwidth]{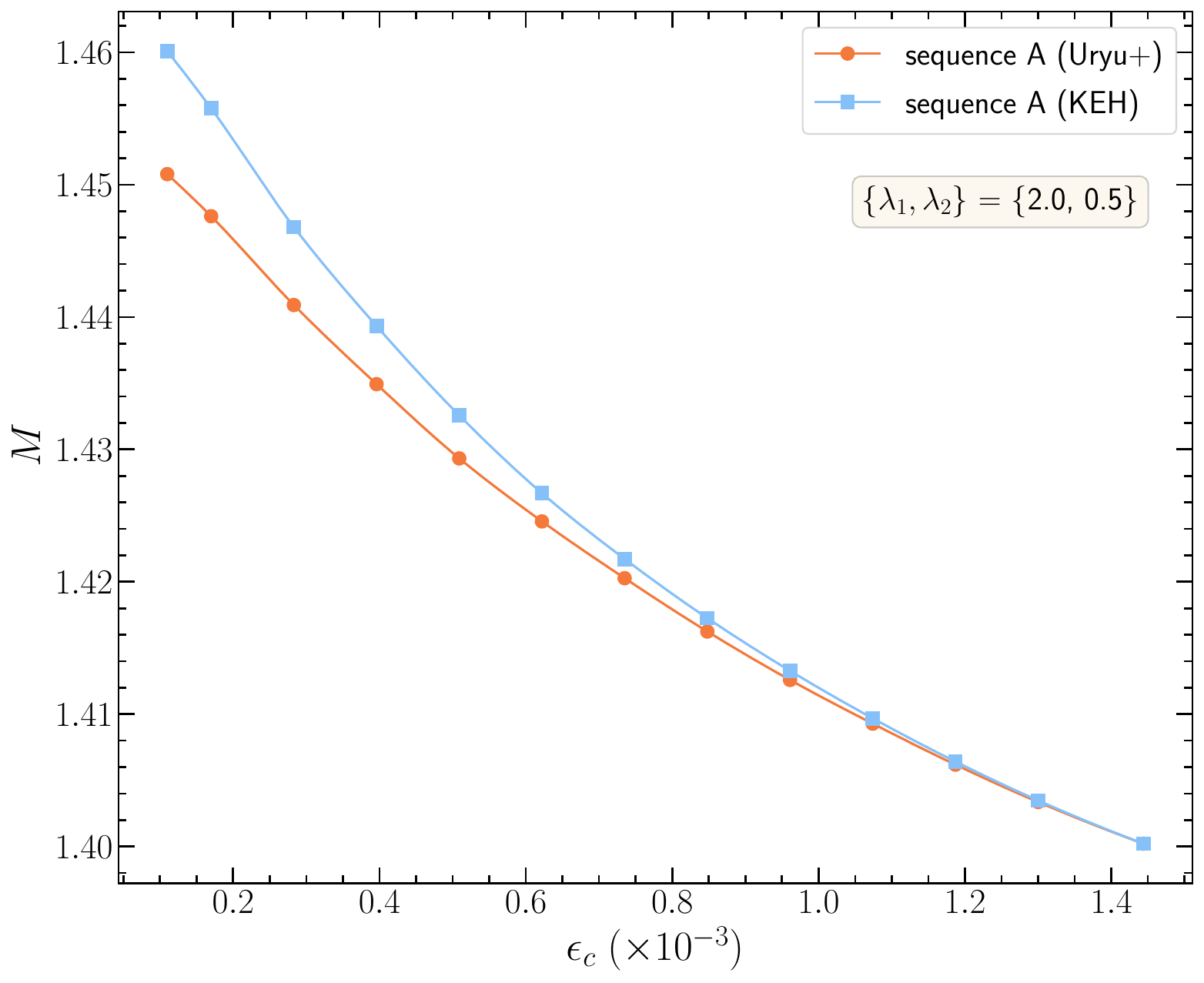}
    \caption{Comparison of gravitational mass $M$ vs. central energy density $\epsilon_c$ for the equilibrium models of sequence A, constructed with the Uryu+ and the KEH differential rotation laws.}
    \label{fig:seqA_mass_Uryu_vs_KEH}
\end{figure}

\begin{figure}
        \includegraphics[width=\columnwidth]{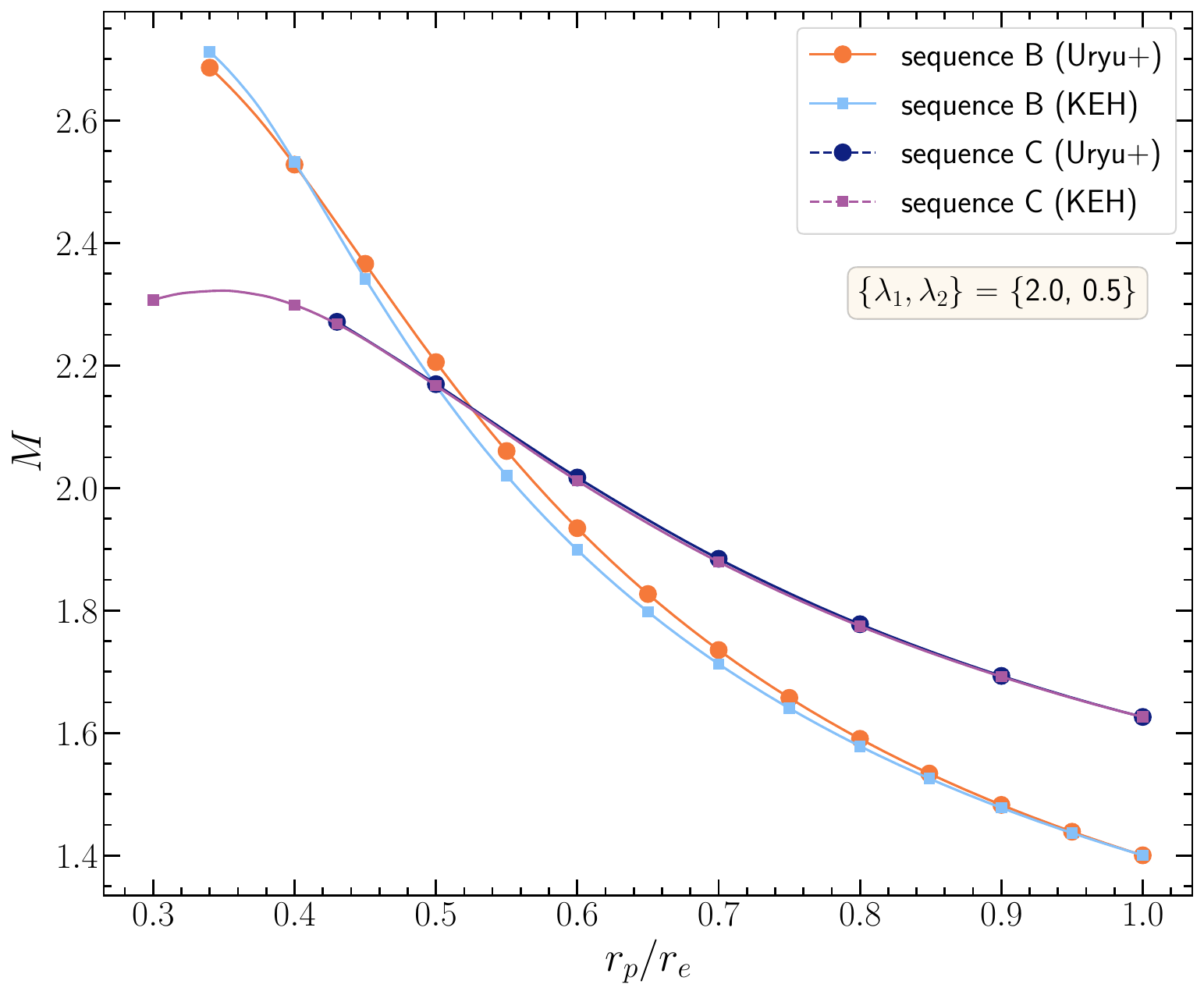}
    \caption{Comparison of gravitational mass $M$ vs. axis ratio $r_p/r_e$ for the equilibrium models of sequences B and C, constructed with the Uryu+ and the KEH differential rotation laws (additional intermediate models were used for sequence C, in order to display a smooth line).}
    \label{fig:seqBC_mass_Uryu_vs_KEH}
\end{figure}

\begin{figure}
        \includegraphics[width=\columnwidth]{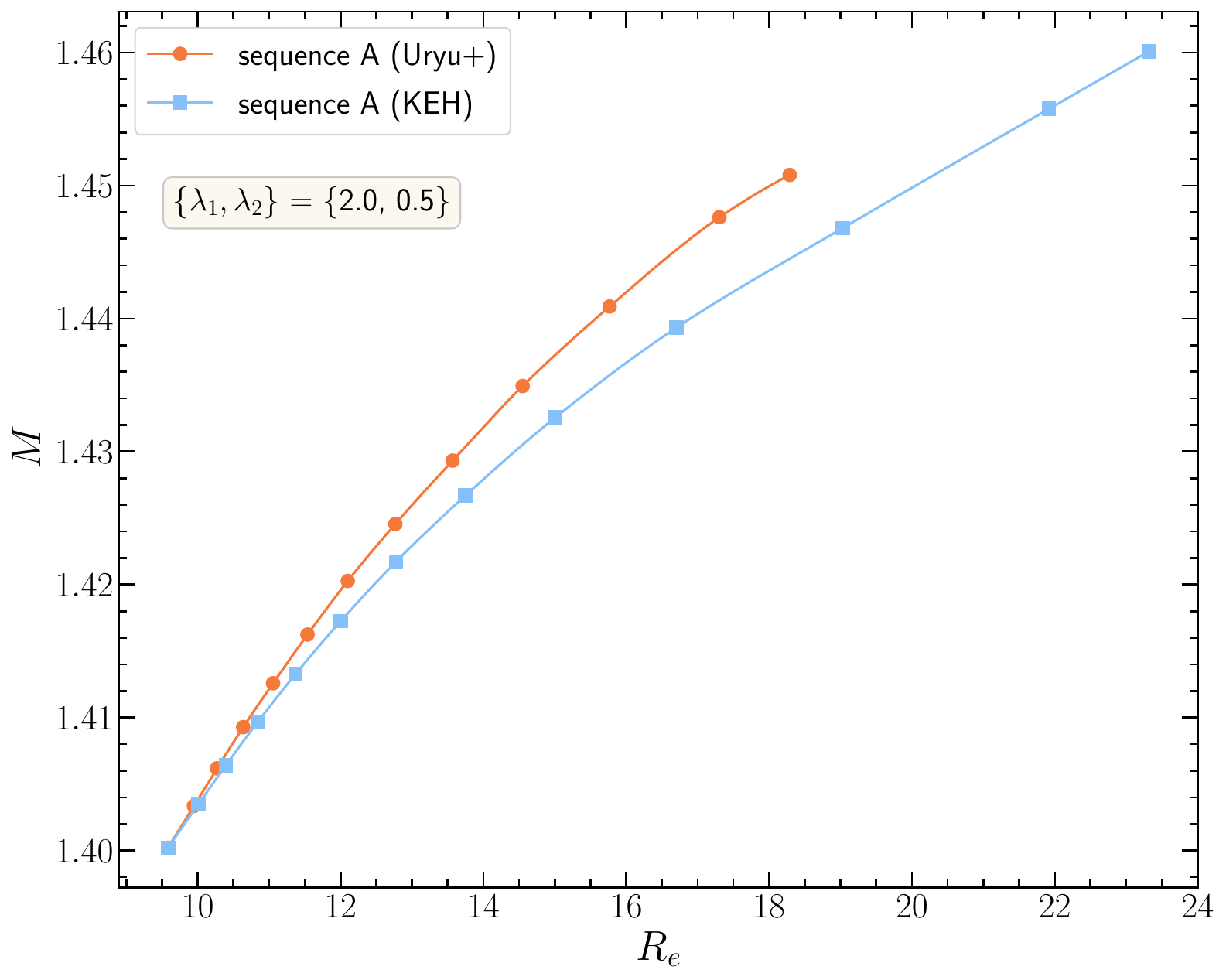}
    \caption{Comparison of gravitational mass $M$ vs. circumferential radius $R_e$ for the equilibrium models of sequence A, constructed with the Uryu+ and the KEH differential rotation laws.}
    \label{fig:seqA_mass_vs_Re_Uryu_vs_KEH}
\end{figure}

\begin{figure}
        \includegraphics[width=\columnwidth]{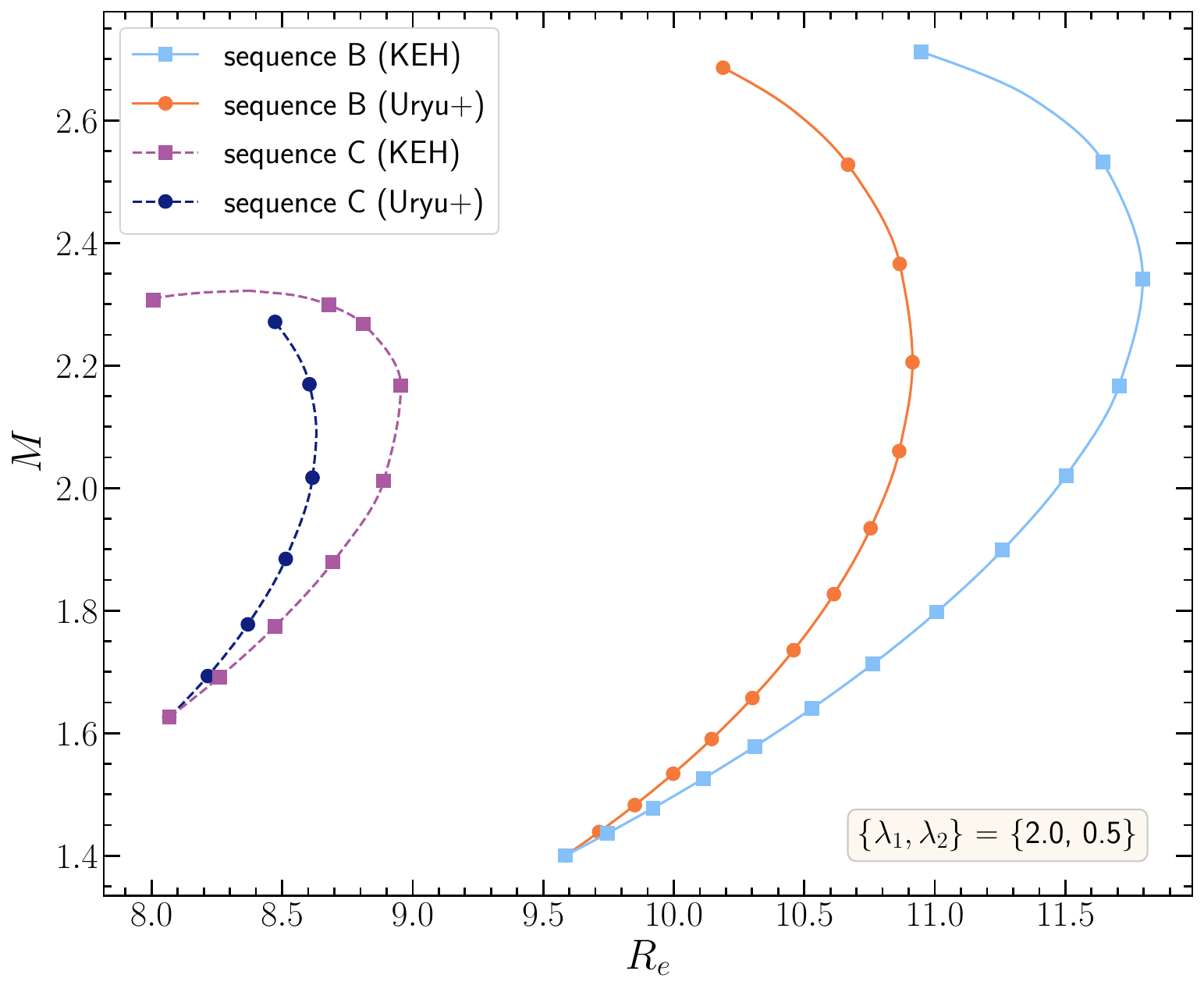}
    \caption{Comparison of gravitational mass $M$ vs. circumferential radius $R_e$ for the equilibrium models of sequences B and C, constructed with the Uryu+ and the KEH differential rotation laws (additional intermediate models were used, in order to display  smooth lines).}
    \label{fig:seqBC_mass_vs_Re_Uryu_vs_KEH}
\end{figure}

Figure~\ref{fig:surface_xz} shows a comparison of the stellar surfaces for representative models of each sequence with an $r_p / r_e$ ratio of $\sim0.5$, between the Uryu+ differential rotation law and the KEH law.  The Uryu+ models have a smaller radius and the surface near the rotation axis has a stronger quasi-toroidal shape than the corresponding KEH models. Note that, {\it based on the density distribution in the equatorial plane},  model A9 would be classified as quasi-spherical for the KEH law with $r_p / r_e = 0.5133$ and marginally quasi-toroidal for the Uryu+ law with $r_p / r_e = 0.4882$, whereas model B10 ($r_p / r_e = 0.5$) would be quasi-spherical for KEH and quasi-toroidal for the Uryu+ law and model C5 ($r_p / r_e = 0.5$) would be classified as quasi-toroidal for both rotation laws. The adoption of the Uryu+ law thus has the tendency to produce quasi-toroidal models earlier along a sequence of equilibrium models that starts at the nonrotating sequence.

Figures~\ref{fig:seqA_mass_Uryu_vs_KEH} and \ref{fig:seqBC_mass_Uryu_vs_KEH} show a comparison of the gravitational mass $M$ along the three equilibrium sequences A, B and C, for models constructed with the Uryu+ and KEH differential rotation laws. Data for the relative physical quantities for the KEH law, appear in \citet{Stergioulas_etal_2004, Iosif_Stergioulas_2014}. We find that the difference in gravitational mass between the two rotation laws is very small, for all three sequences, when one interchanges the two rotation laws.

Figures~\ref{fig:seqA_mass_vs_Re_Uryu_vs_KEH} and \ref{fig:seqBC_mass_vs_Re_Uryu_vs_KEH} compare the mass-radius relations for the three sequences A, B and C, constructed with the two rotation laws. A general remark is that all models using the Uryu+ differential rotation law \eqref{eq:Uryuetal_rotlaw8} have consistently smaller radii than their  counterparts constructed with the KEH law. The relative differences are larger for the rapidly rotating, low-density models of sequence A, where they can exceed $10\%$. For sequence B, the relative differences in radii between the two differential rotation laws, range from $3\%$ to $7\%$ for models with axis ratio between  $r_p / r_e \sim 0.7 $ and $\sim 0.5$, respectively. For sequence C the corresponding relative differences are at the $2\%$ and $4\%$ levels, owing to the greater compactness of the sequence.

\subsection{Variation of parameters in the rotation law}
\label{sec:results}

\begingroup
\setlength{\tabcolsep}{5.7pt}
\begin{table*}
        \centering
        \caption{Comparison of the properties of selected equilibrium models constructed with different choices of $\lambda_1$ and $\lambda_2$, for  models A5, B6 (which have $r_p/r_e\sim 0.7$) and A9, B10 (which have $r_p/r_e\sim 0.5$).}
        \label{tab:lambda12_study}
        \begin{tabular}{lcccccccccccc}
                \hline
                \{$\lambda_1$, $\lambda_2$\} & $\epsilon_\text{max}$ & $M$ & $M_0$ & J & $T/|W|$ & $\Omega_c$ & $\Omega_\text{max}$ & $\Omega_e$ & $\Omega_K$ & $R_e$ & $r_e$ & GRV3\\
                & $\left(\times 10^{-3}\right)$ & & & & $\left(\times 10^{-1}\right)$ & $\left(\times 10^{-2}\right)$ & $\left(\times 10^{-2}\right)$ & $\left(\times 10^{-2}\right)$ & $\left(\times 10^{-2}\right)$ & & & $\left(\times 10^{-5}\right)$\\
                \hline
                A5\\
                \{2.0, 0.5\} & 0.84800 & 1.41624 & 1.50599 & 1.17284 & 0.80274 & 1.97572 & 3.95145 & 0.98786 & 3.08876 & 11.5380 & 10.0283 & 6.47831\\
                \{1.5, 0.5\} & 0.84800 & 1.41101 & 1.49924 & 1.22732 & 0.83749 & 2.14216 & 3.21324 & 1.07108 & 3.00473 & 11.7518 & 10.2458 & 6.51115\\
                \{2.0, 1.0\} & 0.84800 & 1.38326 & 1.46668 & 1.26335 & 0.87340 & 1.26841 & 2.53681 & 1.26841 & 2.79005 & 12.2995 & 10.8216 & 6.56797\\
                \{1.5, 1.0\} & 0.84800 & 1.34909 & 1.42915 & 1.17292 & 0.80248 & 1.44535 & 2.16802 & 1.44535 & 2.67567 & 12.5382 & 11.1014 & 6.58230\\ 
                \\
                A9\\
                \{2.0, 0.5\} & 0.40086 & 1.43493 & 1.50612 & 1.96591 & 1.64514 & 1.97549 & 3.95098 & 0.98775 & 2.28329 & 14.5499 & 12.9843 & 6.03066\\
                \{1.5, 0.5\} & 0.39600 & 1.40917 & 1.47544 & 2.05700 & 1.75397 & 2.04435 & 3.06653 & 1.02218 & 2.11360 & 15.2859 & 13.7453 & 6.06753\\
                \{2.0, 1.0\} & 0.39600 & 1.23786 & 1.28617 & 1.84376 & 1.84378 & 1.04896 & 2.09792 & 1.04896 & 1.60791 & 17.6713 & 16.3254 & 5.72760\\
                \{1.5, 1.0\} & 0.39600 & 0.95152 & 0.98436 & 0.91411 & 1.17554 & 1.07258 & 1.60887 & 1.07258 & 1.16908 & 19.5914 & 18.5953 & 5.17216\\ 
                \\
                B6\\
                \{2.0, 0.5\} & 1.44400 & 1.73532 & 1.88009 & 1.70054 & 0.89312 & 2.76602 & 5.53204 & 1.38301 & 3.91929 & 10.4592 & 8.57816 & 6.38871\\
                \{1.5, 0.5\} & 1.44400 & 1.73978 & 1.88364 & 1.80491 & 0.93770 & 2.97529 & 4.46294 & 1.48765 & 3.79636 & 10.7057 & 8.81697 & 6.49973\\
                \{2.0, 1.0\} & 1.44400 & 1.72067 & 1.85873 & 1.88673 & 0.97554 & 1.72430 & 3.44860 & 1.72430 & 3.48046 & 11.3380 & 9.46891 & 6.63232\\
                \{1.5, 1.0\} & 1.44400 & 1.67444 & 1.80673 & 1.73283 & 0.88100 & 1.93610 & 2.90414 & 1.93610 & 3.30765 & 11.6248 & 9.81449 & 6.71866\\ 
                \\
                B10\\
                \{2.0, 0.5\} & 1.48092 & 2.20535 & 2.41490 & 3.63795 & 1.65695 & 4.10748 & 8.21495 & 2.05374 & 4.15491 & 10.9145 & 8.42730 & 5.64007\\
                \{1.5, 0.5\} & 1.44400 & 2.25934 & 2.47226 & 4.08241 & 1.77762 & 4.19581 & 6.29371 & 2.09790 & 3.85379 & 11.6077 & 9.05178 & 5.50127\\
                \{2.0, 1.0\} & 1.44400 & 2.16091 & 2.34520 & 4.10365 & 1.79888 & 2.03625 & 4.07251 & 2.03625 & 2.92086 & 13.8811 & 11.4650 & 5.75583\\
                \{1.5, 1.0\} & 1.44400 & 1.79857 & 1.94018 & 2.41485 & 1.18892 & 2.03940 & 3.05909 & 2.03940 & 2.27672 & 15.3167 & 13.3840 & 5.82863\\ 
                \\
                \hline
        \end{tabular}
\end{table*}
\endgroup

\begin{figure*}
        \includegraphics[scale=0.349]{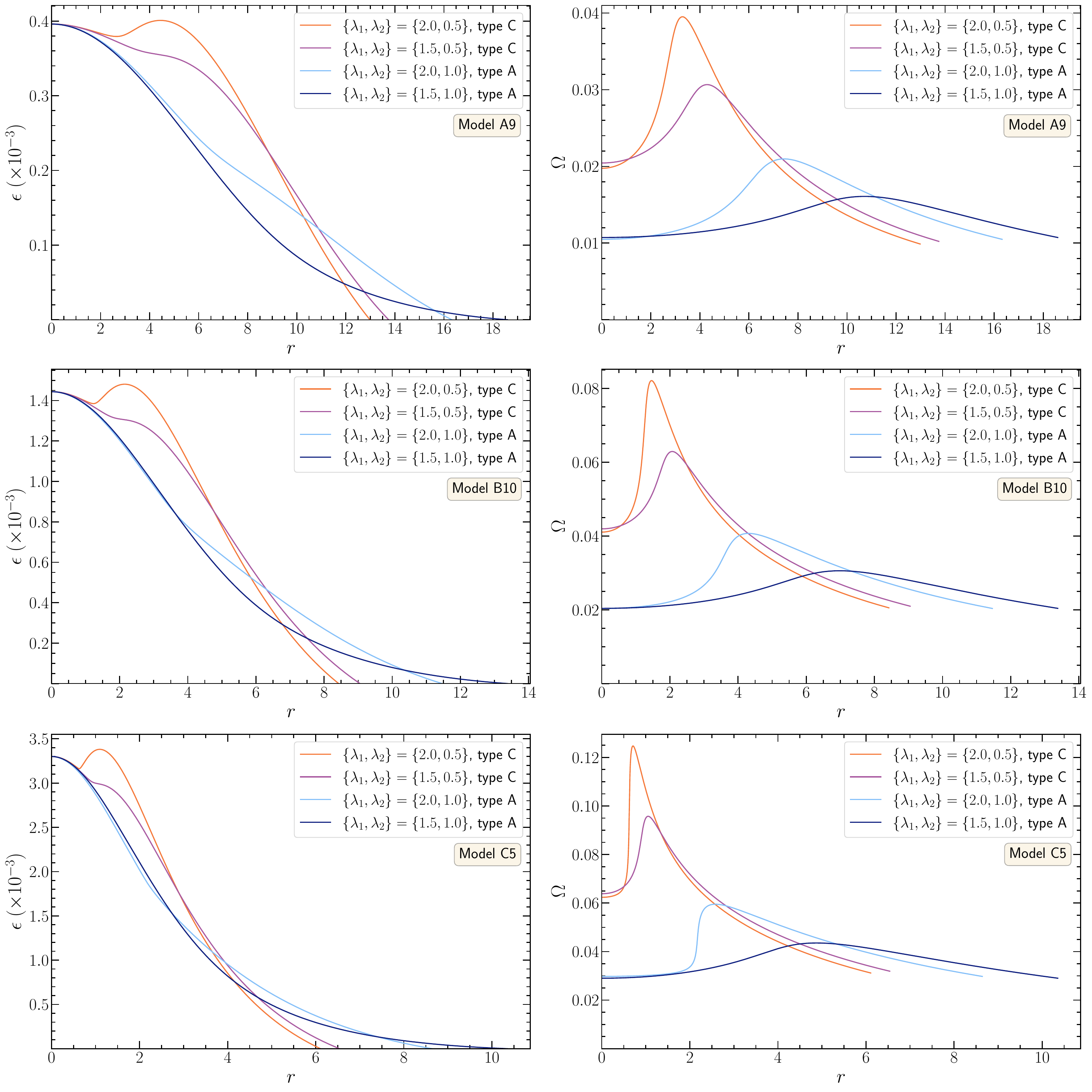}
    \caption{{\it Left column}: Effect of different options for  $\{\lambda_1, \lambda_2\}$ on the energy density profile $\epsilon(r)$ versus the coordinate radius $r$, in the equatorial plane for models A9 ($ r_p/r_e = 0.4882$), B10 ($ r_p/r_e = 0.5$) and C5 ($ r_p/r_e = 0.5$). {\it Right column}: Same as left column, but for the angular velocity profile $\Omega(r)$ in the equatorial plane.}
    \label{fig:diff_lambda_options}
\end{figure*}

The models defined in Sec. \ref{sec:definitions} and
explored in Sec. \ref{sec:comparisons} were constructed with the specific choice of $p=1, q=3$ and $\lambda_1 = 2.0, \lambda_2 = 0.5$. Here, we explore the impact of varying $\lambda_1$ and $\lambda_2$, while keeping $p=1, q=3$ fixed, by first focusing on selected equilibrium models from each sequence with $r_p/r_e \sim 0.7 $ and $\sim 0.5$. Table~\ref{tab:lambda12_study} summarises our results for representative models of sequences A and B with axis ratio values of 0.7 and 0.5, while full sequence C variations are explored in Table~\ref{tab:seqC_variants}. We pick two values for the first parameter $\lambda_1 = \{2.0, 1.5\}$ and two values for the second parameter $\lambda_2 = \{0.5, 1.0\}$, leading to four distinct pairs of $(\lambda_1, \lambda_2)$. This is motivated by the differences in rotational profiles seen in numerical simulations of post-merger remnants, when different EOS and total masses are used, see the examples in \citet{Hanauske_etal_2017, DePietri_etal_2020}. 

Models A5, B6 in Table~\ref{tab:lambda12_study}, and C3 in Table~\ref{tab:seqC_variants} have $r_p / r_e \sim 0.7$. Between the different $\{\lambda_1, \lambda_2 \}$ runs, we notice variations in all the equilibrium quantities. The most striking difference is the fact that the equatorial radii are increasing with decreasing $\lambda_1$ and increasing  $\lambda_2$. Increasing $\lambda_2=\Omega_e / \Omega_c$ leads to a larger radius increase, compared to the corresponding increase in radius when  $\lambda_1=\Omega_\text{max} / \Omega_c$ decreases.

Models A9, B10 in Table~\ref{tab:lambda12_study} and C5 in Table~\ref{tab:seqC_variants} have $r_p / r_e \sim 0.5$. The increase of the equatorial radii with 
decreasing $\lambda_1$ and with increasing $\lambda_2$ is also present here, with the effect magnified due to the faster rotation. Again, increasing $\lambda_2=\Omega_e / \Omega_c$ leads to a larger radius increase, compared to the corresponding increase in radius when  $\lambda_1=\Omega_\text{max} / \Omega_c$ decreases. 
The difference in masses (both gravitational and rest mass) between the different $\{\lambda_1, \lambda_2 \}$ runs, is more pronounced in the low-density configuration A9 compared to the other two models B10 and C5 that are more compact. An important observation is that while these models are quasi-toroidal for the choice $\{\lambda_1, \lambda_2 \} = \{ 2.0, 0.5\}$, they are quasi-spherical for the other three $\{\lambda_1, \lambda_2 \}$ options we consider here (since $\epsilon_c = \epsilon_\text{max}$).

In Figure~\ref{fig:diff_lambda_options} we present the effect of the different $\lambda_1$ and $\lambda_2$ options on the energy density and angular velocity profiles in the equatorial plane, for the three models A9, B10 and C5 with $r_p/r_e \sim 0.5$.\footnote{Depending on the cut-off density, the EOS and the total mass, the axis ratio $r_p/r_e$  of post-merger remnants in simulations is roughly in the range between $\sim 0.5$ and $\sim 0.7$.}. The figures in the left column of Figure~\ref{fig:diff_lambda_options} show the energy density profile $\epsilon(r)$ for the four different $\{\lambda_1, \lambda_2 \}$ pairs considered. As we already pointed out, apart from the default option $\{ \lambda_1, \lambda_2 \} = \{2.0, 0.5\}$ that yields quasi-toroidal configurations for models A9, B10 and C5, the other three $\{ \lambda_1, \lambda_2 \}$ pairs result in quasi-spherical configurations for the same $\epsilon_c$ and $r_p / r_e$ values. The figures in the right column of Figure~\ref{fig:diff_lambda_options} show the corresponding angular velocity profile $ \Omega(r)$  for the four distinct $\{\lambda_1, \lambda_2 \}$ pairs for each of the three chosen models, demonstrating in essence the different degrees of differential rotation considered. We notice that any choice other than the default $\{ \lambda_1, \lambda_2 \} = \{2.0, 0.5\}$, results in larger radii, with respect to this choice. 

In order to probe whether the morphology of the selected models A9, B10 and C5 falls into the spheroidal or quasi-toroidal category, we kept each model's central energy density fixed and constructed additional configurations at lower axis ratios for each of the four pairs $\{ \lambda_1, \lambda_2 \}$ under study. For the three values of central energy density considered (i.e. model A9's $\epsilon_c = 0.396 \times 10^{-3}$, model B10's $\epsilon_c = 1.444 \times 10^{-3}$ and model C5's $\epsilon_c = 3.3 \times 10^{-3}$) we found that both $\{ \lambda_1, \lambda_2 \}$ variations with $\lambda_2 = 0.5$ (i.e. the pairs $\{2.0, 0.5\}$ and $\{1.5, 0.5\}$) exhibited strong quasi-toroidal characteristics at axis ratios lower than 0.5. Respectively, both $\{ \lambda_1, \lambda_2 \}$ variants with $\lambda_2 = 1.0$ (i.e. the pairs $\{2.0, 1.0\}$ and $\{1.5, 1.0\}$) came gradually closer to the mass-shedding limit, as the axis ratio was reduced below 0.5, while remaining quasi-spherical. This behaviour is reminiscent of different types of solutions in the \citet{Ansorg_etal_2009} classification scheme (where four different types of solutions, A, B, C and D were identified) and encourages a more thorough investigation of variants of complete sequences of models with different $\{ \lambda_1, \lambda_2 \}$ options, which we provide in the following Section.

\subsection{Identification of type A and C solutions}
\label{sec:solutions_classification}

\begingroup
\setlength{\tabcolsep}{3.0pt}
\begin{table*}
        \centering
        \caption{Physical quantities for variations of sequence C (with constant  $\epsilon_c = 3.3 \times 10^{-3}$) calculated using the Uryu+ differential rotation law \eqref{eq:Uryuetal_rotlaw8} and four different $\{\lambda_1, \lambda_2\}$ choices. The different quantities are defined as in Table \ref{tab:convergence_study}.
        } 
       
        \label{tab:seqC_variants}
        \begin{tabular}{lcccccccccccccccc}
                \hline
                Model & $r_p/r_e$ & $\epsilon_c$ & $\epsilon_\text{max}$ & $M$ & $M_0$ & J & $T/|W|$ & $\Omega_c$ & $\Omega_\text{max}$ & $\Omega_e$ & $\Omega_K$ & $R_e$ & $r_e$ & GRV3\\
                $\{\lambda_1, \lambda_2\}$ & & $\left(\times 10^{-3}\right)$ & $\left(\times 10^{-3}\right)$ & & & & $\left(\times 10^{-1}\right)$ & $\left(\times 10^{-2}\right)$ & $\left(\times 10^{-2}\right)$ & $\left(\times 10^{-2}\right)$ & $\left(\times 10^{-2}\right)$ & & & $\left(\times 10^{-5}\right)$\\
                \hline
                C0 & 1.0 & 3.300 & 3.30000 & 1.62609 & 1.78404 & 0.0 & 0.0 & 0.0 & 0.0 & 0.0 & 5.56489 & 8.06761 & 6.33721 & 4.07290\\ \\
                $\{2.0, 0.5\}$\\
                C1 & 0.900 & 3.300 & 3.30000 & 1.69308 & 1.85770 & 0.82658 & 0.26133 & 2.21234 & 4.42469 & 1.10617 & 5.45745 & 8.21539 & 6.39361 & 8.22374\\
                C2 & 0.800 & 3.300 & 3.30000 & 1.77749 & 1.95056 & 1.33108 & 0.55526 & 3.24900 & 6.49800 & 1.62450 & 5.43431 & 8.36885 & 6.43192 & 7.27867\\
                C3 & 0.700 & 3.300 & 3.30000 & 1.88427 & 2.06789 & 1.88720 & 0.88511 & 4.16835 & 8.33671 & 2.08418 & 5.45131 & 8.51392 & 6.43013 & 6.50502\\
                C4 & 0.600 & 3.300 & 3.30000 & 2.01691 & 2.21281 & 2.55894 & 1.25159 & 5.11713 & 10.2343 & 2.55856 & 5.53276 & 8.61636 & 6.34583 & 5.92909\\
                C5 & 0.500 & 3.300 & 3.38103 & 2.16935 & 2.37578 & 3.36450 & 1.64806 & 6.23604 & 12.4721 & 3.11802 & 5.73080 & 8.60458 & 6.10795 & 5.50829\\
                C6 & 0.430 & 3.300 & 3.78169 & 2.27114 & 2.47788 & 3.95736 & 1.93348 & 7.22363 & 14.4473 & 3.61181 & 5.97797 & 8.47292 & 5.81060 & 5.31201\\ 
                \\
                $\{1.5, 0.5\}$\\
                C1 & 0.900 & 3.300 & 3.30000 & 1.69744 & 1.86262 & 0.87469 & 0.27516 & 2.43702 & 3.65552 & 1.21851 & 5.41157 & 8.26711 & 6.43940 & 8.41869\\
                C2 & 0.800 & 3.300 & 3.30000 & 1.78925 & 1.96400 & 1.42246 & 0.58641 & 3.53712 & 5.30569 & 1.76856 & 5.33519 & 8.49234 & 6.53988 & 7.63975\\
                C3 & 0.700 & 3.300 & 3.30000 & 1.90963 & 2.09728 & 2.04905 & 0.93923 & 4.46770 & 6.70156 & 2.23385 & 5.28216 & 8.74006 & 6.62358 & 6.92976\\
                C4 & 0.600 & 3.300 & 3.30000 & 2.06929 & 2.27439 & 2.85553 & 1.33838 & 5.37174 & 8.05761 & 2.68587 & 5.26961 & 8.99051 & 6.65339 & 6.26718\\
                C5 & 0.500 & 3.300 & 3.30000 & 2.27675 & 2.50393 & 3.93817 & 1.78309 & 6.38320 & 9.57480 & 3.19160 & 5.35024 & 9.17495 & 6.54069 & 5.65454\\
                C6 & 0.450 & 3.300 & 3.30000 & 2.39419 & 2.63214 & 4.58693 & 2.01748 & 6.99955 & 10.4993 & 3.49977 & 5.45900 & 9.19236 & 6.38077 & 5.39346\\
                C7 & 0.400 & 3.300 & 3.37195 & 2.50860 & 2.75322 & 5.25894 & 2.25273 & 7.74559 & 11.6184 & 3.87280 & 5.63695 & 9.12106 & 6.12492 & 5.19915\\
                C8 & 0.350 & 3.300 & 3.78944 & 2.60300 & 2.84493 & 5.86720 & 2.48126 & 8.65126 & 12.9769 & 4.32563 & 5.88738 & 8.95023 & 5.78398 & 5.11267\\
                C9 & 0.310 & 3.300 & 4.25504 & 2.65527 & 2.88547 & 6.25344 & 2.65580 & 9.47333 & 14.2101 & 4.73666 & 6.11700 & 8.76773 & 5.49073 & 5.14309\\
                \\
                $\{2.0, 1.0\}$\\
                C1 & 0.900 & 3.300 & 3.30000 & 1.70172 & 1.86729 & 0.93253 & 0.29149 & 1.49441 & 2.98881 & 1.49441 & 5.29985 & 8.39149 & 6.55728 & 8.57625\\
                C2 & 0.800 & 3.300 & 3.30000 & 1.79926 & 1.97507 & 1.52450 & 0.61561 & 2.10463 & 4.20927 & 2.10463 & 5.08468 & 8.79815 & 6.83195 & 7.86928\\
                C3 & 0.700 & 3.300 & 3.30000 & 1.92722 & 2.11714 & 2.20949 & 0.97340 & 2.54465 & 5.08929 & 2.54465 & 4.83450 & 9.32979 & 7.19275 & 7.22397\\
                C4 & 0.600 & 3.300 & 3.30000& 2.09626 & 2.30584 & 3.09888 & 1.35888 & 2.85349 & 5.70699 & 2.85349 & 4.51125 & 10.0721 & 7.71297 & 6.72010\\
                C5 & 0.500 & 3.300 & 3.30000 & 2.30708 & 2.54232 & 4.26788 & 1.73331 &  2.97673 & 5.95346 & 2.97673 & 4.01493 & 11.2684 & 8.64295 & 6.30155\\
                C6 & 0.450 & 3.300 & 3.30000 & 2.39295 & 2.63801 & 4.80388 & 1.84621 & 2.88492 & 5.76984 & 2.88492 & 3.58163 & 12.3292 & 9.61201 & 6.08501\\
                C7 & 0.400 & 3.300 & 3.30000 & 2.22585 & 2.44598 & 3.95290 & 1.53164 & 2.47488 & 4.94977 & 2.47488 & 2.68011 & 14.6073 & 12.1649 & 6.28917\\
                C8 & 0.38602 & 3.300 & 3.30000 & 2.02169 & 2.21845 & 2.79739 & 1.13836 & 2.26348 & 4.52696 & 2.26348 & 2.26353 & 15.7686 & 13.6103 & 6.60086\\
                \\
                $\{1.5, 1.0\}$\\
                C1 & 0.900 & 3.300 & 3.30000 & 1.69717 & 1.86235 & 0.91215 & 0.27301 & 1.71448 & 2.57171 & 1.71448 & 5.24254 & 8.44636 & 6.61807 & 8.63909\\
                C2 & 0.800 & 3.300 & 3.30000 & 1.78461 & 1.95901 & 1.46353 & 0.56461 & 2.37684 & 3.56526 & 2.37684 & 4.94888 & 8.93946 & 6.99342 & 8.02860\\
                C3 & 0.700 & 3.300 & 3.30000 & 1.88878 & 2.07452 & 2.04385 & 0.86078 & 2.80492 & 4.20738 & 2.80492 & 4.57924 & 9.61981 & 7.53786 & 7.47561\\
                C4 & 0.600 & 3.300 & 3.30000 & 1.99392 & 2.19113 & 2.62218 & 1.11124 & 3.01744 & 4.52616 & 3.01744 & 4.05461 & 10.6529 & 8.44373 & 7.01096\\
                C5 & 0.500 & 3.300 & 3.30000 & 1.99260 & 2.18826 & 2.64193 & 1.08973 & 2.90352 & 4.35528 & 2.90352 & 3.18093 & 12.5257 & 10.3560 & 6.79391\\
                C6 & 0.490 & 3.300 & 3.30000 & 1.97350 & 2.16693 & 2.53802 & 1.04493 & 2.86533 & 4.29799 & 2.86533 & 3.06002 & 12.8046 & 10.6660 & 6.79974\\
                C7 & 0.480 & 3.300 & 3.30000 & 1.94882 & 2.13949 & 2.40167 & 0.98693 & 2.82147 & 4.23221 & 2.82147 & 2.93199 & 13.1080 & 11.0081 & 6.83711\\
                C8 & 0.46693 & 3.300 & 3.30000 & 1.90861 & 2.09501 & 2.17639 & 0.89089 & 2.75619 & 4.13429 & 2.75619 & 2.75620 & 13.5436 & 11.5046 & 6.86623\\
                \hline
        \end{tabular}
\end{table*}
\endgroup

\begin{figure*}
    \includegraphics[scale=0.43]{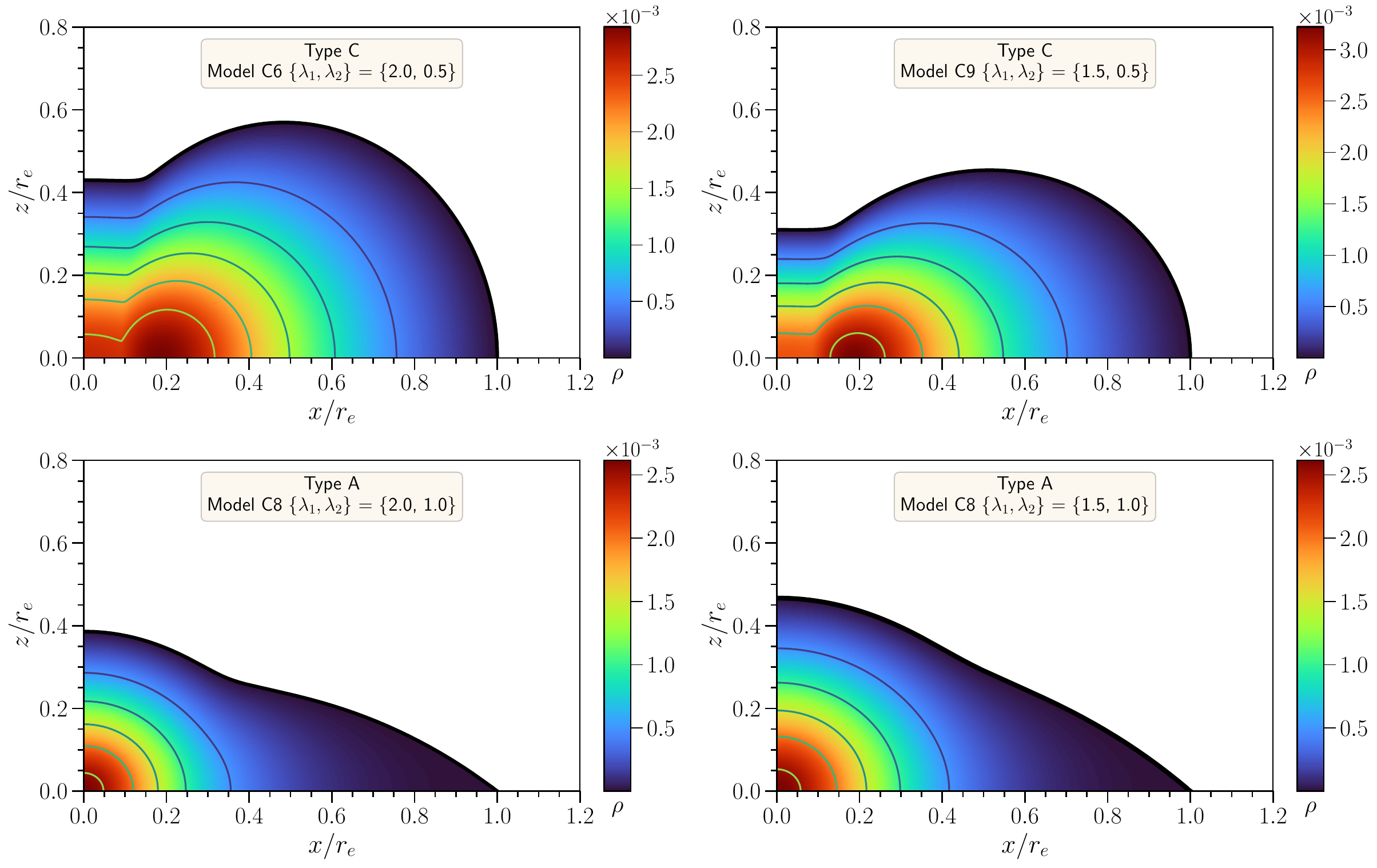}
    \caption{Two-dimensional rest mass density distribution $\rho$ (in dimensionless units $c = G = M_\odot = 1 $) for the terminal models of each variation of sequence C for the different $\{\lambda_1, \lambda_2\}$ pairs employed (see Table~\ref{tab:seqC_variants}). The solid black curves represent the models' surfaces. The coordinates $x$ and $z$ are rescaled with each model's coordinate radius at the equator $r_e$. The mass-shedding limit is encountered for $\{\lambda_1, \lambda_2\} = \{2.0, 1.0\}$ at $r_p/r_e = 0.38602$ and for $\{\lambda_1, \lambda_2\} = \{1.5, 1.0\}$ at $r_p/r_e = 0.46693$, meaning sequence C belongs to type A solutions \citep{Ansorg_etal_2009} for these choices of parameters. No mass-shedding limit is found for $\{\lambda_1, \lambda_2\} = \{2.0, 0.5\}$, where the terminal model C6 is reached at $r_p/r_e=0.43$ and $\{\lambda_1, \lambda_2\} = \{1.5, 0.5\}$, where the terminal model C9 is reached at $r_p/r_e=0.31$. The appearance of quasi-toroidal models characterizes these variations of sequence C as type C solutions according to \citeauthor{Ansorg_etal_2009}.}
    \label{fig:seqC_variations_density2Dxz}
\end{figure*}

\begin{figure*}
    \includegraphics[scale=0.57]{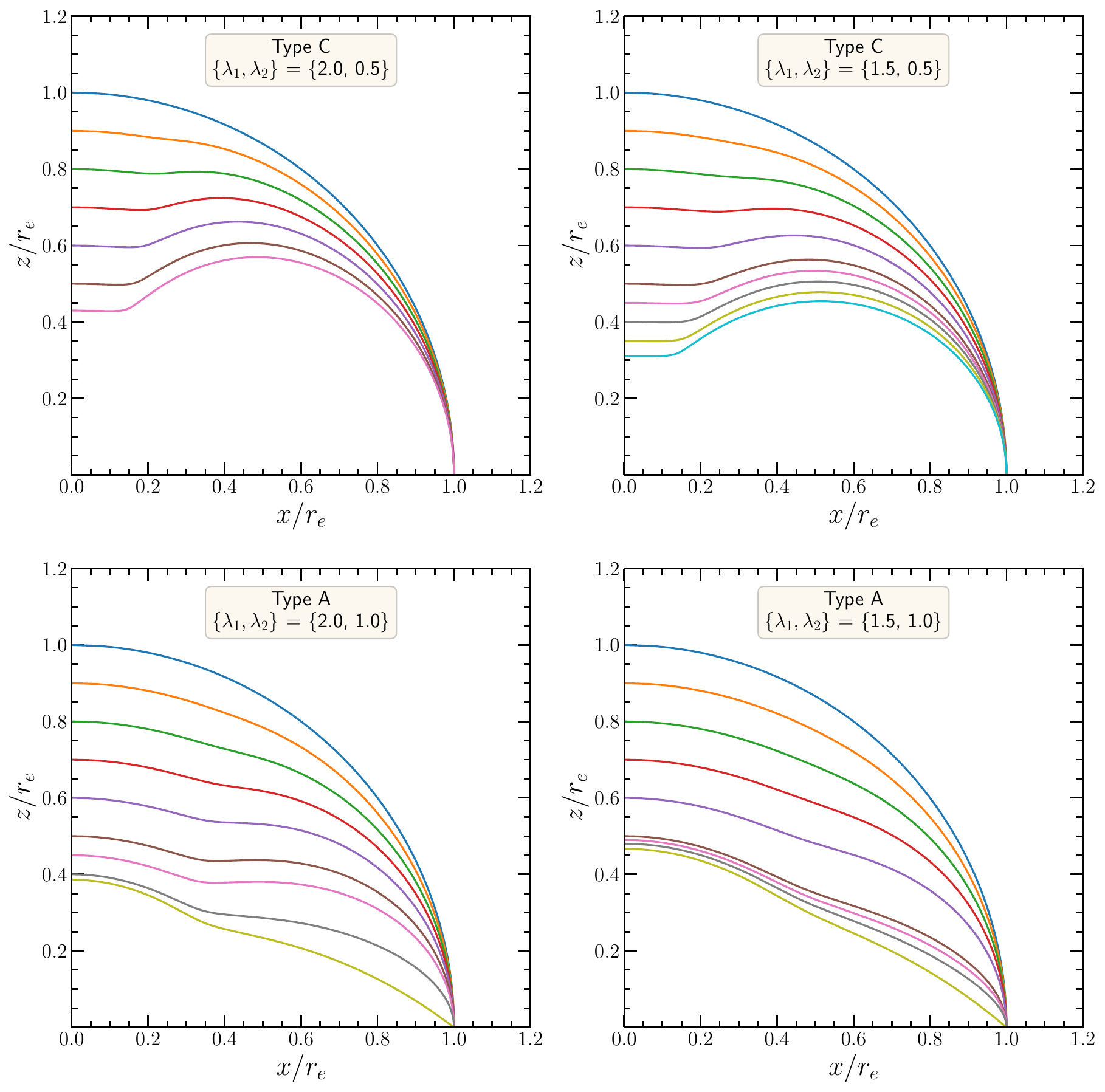}
    \caption{Surface shape for all variations of sequence C for the different $\{\lambda_1, \lambda_2\}$ pairs employed (see Table~\ref{tab:seqC_variants}). The coordinates $x$ and $z$ are rescaled with each model's coordinate radius at the equator $r_e$. The different colors in each panel represent the different models of each variation of sequence C, as reported in Table~\ref{tab:seqC_variants}. The upper panels represent type C solutions according to the classification of \citet{Ansorg_etal_2009} for values $\{\lambda_1, \lambda_2\} = \{2.0, 0.5\}$ and $\{\lambda_1, \lambda_2\} = \{1.5, 0.5\}$, while the lower panels represent type A solutions for values $\{\lambda_1, \lambda_2\} = \{2.0, 1.0\}$ and $\{\lambda_1, \lambda_2\} = \{1.5, 1.0\}$.}
    \label{fig:seqC_variations_surfaces}
\end{figure*}

Having performed an initial exploration for selected equilibrium models, of the effect that different values for the parameters $\{\lambda_1, \lambda_2\}$ have, we now construct variants of the complete sequence C, employing four different pairs of $\{\lambda_1, \lambda_2\}$. Our results for the four variants are reported in Table~\ref{tab:seqC_variants}. 

For the cases of $\{\lambda_1, \lambda_2\} = \{1.5, 1.0\}$ and $\{2.0, 1.0\}$ we find that the sequence reaches the mass-shedding limit (where $\Omega_e = \Omega_K$) at an axis ratio $r_p/r_e$ of 0.46693 and 0.38602, correspondingly. From Table~\ref{tab:seqC_variants}, we notice that the these models at the mass-shedding limit are still quasi-spherical ($\epsilon_c = \epsilon_\text{{max}}$). The lower panels of Figure~\ref{fig:seqC_variations_density2Dxz} show that the two-dimensional rest-mass density distributions of these two models at the mass-shedding limit feature the characteristic cusp in the equatorial plane. In addition, the lower panels of Figure~\ref{fig:seqC_variations_surfaces} show the shape of the surface for all members of the two variants of sequence C that terminate at the mass-shedding limit. Evidently, these two variants of sequence C remain quasi-spherical, up to mass-shedding. We conclude that \textit{the variants of sequence C calculated with $\{\lambda_1, \lambda_2\} = \{1.5, 1.0\}$ and $\{\lambda_1, \lambda_2\} = \{2.0, 1.0\}$ can be classified as type A solutions\footnote{We caution that our terminology of "sequence A" should not be confused with the terminology of "type A solutions". }, according to the classification of \citet{Ansorg_etal_2009}.}

For the variant of sequence C with $\{\lambda_1, \lambda_2\} = \{2.0, 0.5\}$, quasi-toroidal models appear already at $r_p /r_e = 0.5$ (see Table~\ref{tab:seqC_variants}, where for model C5 of this variation, $ \epsilon_c < \epsilon_\text{{max}} $, i.e. the maximum density is off-center). The upper left panel of Figure~\ref{fig:seqC_variations_density2Dxz} shows the meridional rest mass density distribution for the terminal model of this variation (model C6 with $r_p/r_e = 0.43$). The corresponding panel of Figure~\ref{fig:seqC_variations_surfaces} shows that the shape of the surface for the members of this sequence is transitioning from quasi-spheroidal models at high axis ratio to quasi-toroidal models at low axis ratio.
Since the mass-shedding limit has not been reached (see Table~\ref{tab:seqC_variants}), even though the models have become quasi-toroidal, we classify the models belonging to this variant of sequence C as type C solutions\footnote{We caution that our terminology of "sequence C" should not be confused with the terminology of "type C solutions".}, following \citet{Ansorg_etal_2009}.

For the variant of sequence C with $\{\lambda_1, \lambda_2\} = \{1.5, 0.5\}$ the situation is similar to the case of  $\{\lambda_1, \lambda_2\} = \{2.0, 0.5\}$. After constructing the initial quasi-spheroidal models of this family (i.e. models with $r_p/r_e > 0.5$ where $\epsilon_c = \epsilon_\text{{max}}$, see Table~\ref{tab:seqC_variants}) we performed a search starting from an axis ratio value of 0.5 and decreased $r_p /r_e$ using a 0.01 step. We noticed a local maximum in the energy density profile at the equatorial plane gradually rising and when the axis ratio reached the value 0.41, the maximum density was found to be definitively off-center. This marked the transition to quasi-toroidal configurations. We kept on decreasing the axis ratio, down to a value of 0.31, where the {\tt rns} code ceased producing models for this choice of parameters. The upper right panel of Figure~\ref{fig:seqC_variations_density2Dxz} shows the meridional rest mass density distribution for the terminal model C9 (with $ r_p / r_e = 0.31 $). Models constructed with $0.41 > r_p / r_e \geq 0.31$ exhibited a gradually “stronger” quasi-toroidal morphology (see upper right panel of Figure~\ref{fig:seqC_variations_surfaces}), while $\Omega_e$ always remained smaller than $\Omega_K$. We note that for the terminal model C9, we have $\Omega_e / \Omega_K \simeq 0.77$, i.e. no mass-shedding limit is encountered. Thus, we classify \textit{the variant of sequence C, calculated with $\{\lambda_1, \lambda_2\} = \{1.5, 0.5\}$, as a type C solution, according to \citet{Ansorg_etal_2009}.}

We note, that as can be seen from the right column of Figure~\ref{fig:diff_lambda_options}, the pairs $\{\lambda_1, \lambda_2\} = \{1.5, 1.0\}$ and $\{\lambda_1, \lambda_2\} = \{2.0, 1.0\}$ that yield type A solutions represent lower degrees of differential rotation compared to the pairs $\{\lambda_1, \lambda_2\} = \{2.0, 0.5\}$ and $\{\lambda_1, \lambda_2\} = \{1.5, 0.5\}$ that result in type C solutions. The pair $\{\lambda_1, \lambda_2\} =\{2.0, 0.5\}$ corresponds to the highest degree of differential rotation and the pair $\{1.5, 1.0\}$ to the lowest. This observation together with the classification of solutions described above are in agreement with \citet{Ansorg_etal_2009}, where it was demonstrated that type C solutions arise for a larger degree of differential rotation, whereas type A solutions correspond to a lower degree of differential rotation.

\begin{figure}
        \includegraphics[width=\columnwidth]{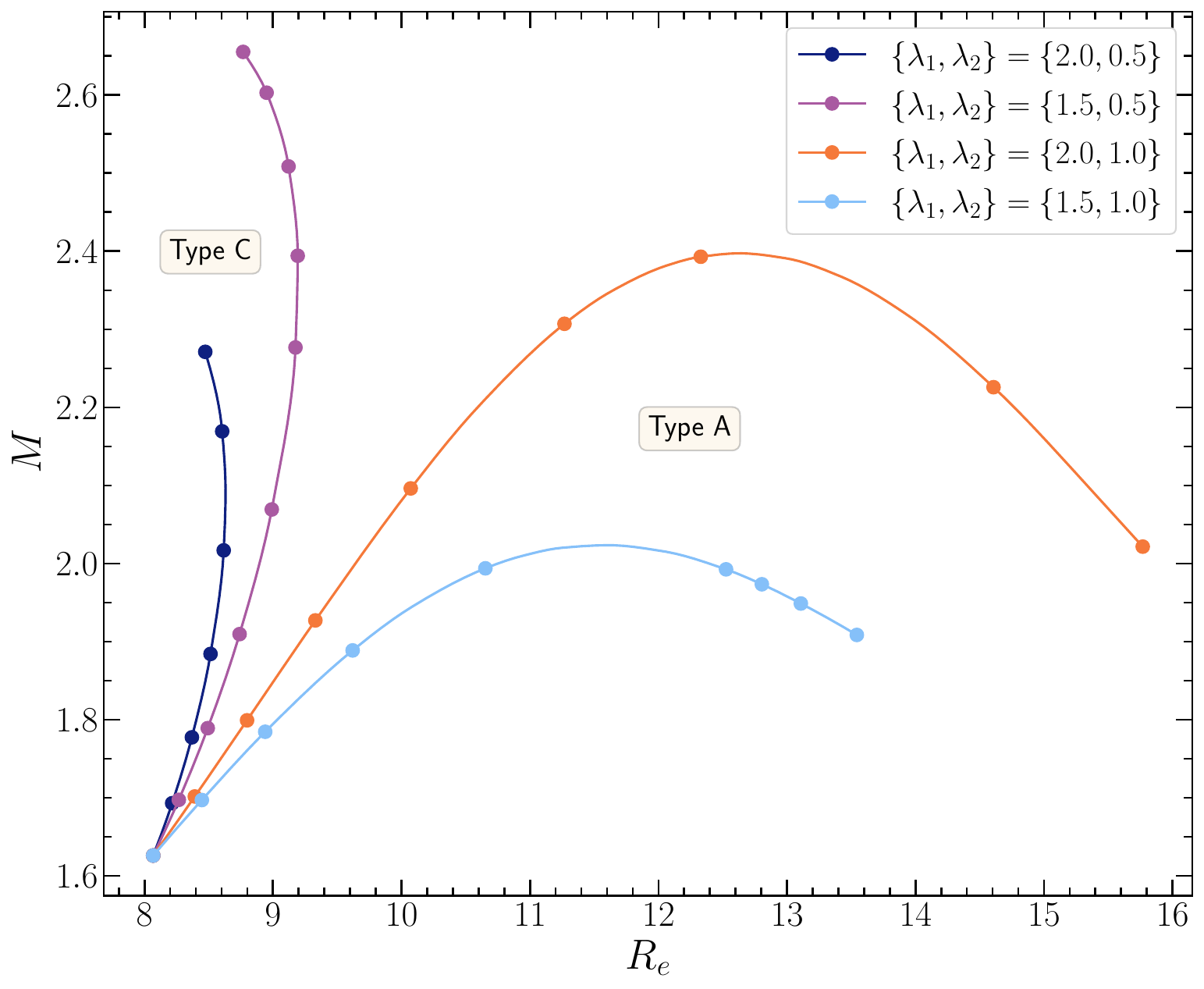}
    \caption{Comparison of gravitational mass $M$ vs. circumferential radius $R_e$ for the variations of sequence C, constructed with the Uryu+ differential rotation law and employing different $\{\lambda_1, \lambda_2\}$ values (see Table~\ref{tab:seqC_variants}). Equilibrium models with $\{\lambda_1, \lambda_2\} = \{2.0, 0.5\}$ and $\{1.5, 0.5\}$ are type C solutions, while models with $\{\lambda_1, \lambda_2\} = \{2.0, 1.0\}$ and $\{1.5, 1.0\}$ are type A solutions according to \citet{Ansorg_etal_2009} (additional intermediate models were used, in order to display  smooth lines).}
    \label{fig:seqC_variations_mass_vs_Re}
\end{figure}

Finally, Figure~\ref{fig:seqC_variations_mass_vs_Re} further demonstrates the distinct types of solutions for the choices of $\{\lambda_1, \lambda_2\}$ considered. In the mass vs. equatorial radius plane, the four sequence C variants form a "bouquet" of equilibrium sequences where type A solutions are clearly distinguished from type C solutions. The sequences with type A solutions have a monotonically increasing radius and exhibit a maximum mass. In contrast, the sequences with type C solutions have a monotonically increasing mass and exhibit a maximum radius (which is significantly smaller than the radius of most type A models). A more detailed investigation of sequences with different $\{\lambda_1, \lambda_2\}$ values may reveal the location of a separatrix in between the two types of solutions. 

We note that the variation of $\lambda_2$  appears to have the strongest effect regarding the type of solutions: keeping $\lambda_1$ constant and varying $\lambda_2$ from 0.5 to 1.0, results to a transition from type C to type A solutions (observe Figure~\ref{fig:seqC_variations_surfaces} per column of panels), whereas if we keep $\lambda_2$ fixed and vary $\lambda_1$ from 2.0 to 1.5 the type of solution remains the same (observe Figure~\ref{fig:seqC_variations_surfaces} per row of panels). The above observations hold for the particular values of $\{p,q\} = \{1,3\}$ of the Uryu+ rotation law \eqref{eq:Uryuetal_rotlaw8} that are kept fixed within the scope of this work.

\section{Summary and Conclusions}
\label{sec:conclusions}

In this work, our goal was to investigate the new 4-parameter differential rotation law \eqref{eq:Uryuetal_rotlaw8} introduced in \citet{Uryu_etal_2017}, as a more realistic option compared to the classic $j$-constant law \eqref{eq:KEH_rotlaw} for describing compact remnants from BNS mergers, in quasi-equilibrium. We did this by constructing sequences of equilibrium models, that can be used as reference models to compare with different rotation law implementations. We found that the new 4-parameter rotation law provides much more versatility as a candidate for equilibrium model building. The Uryu+ law constitutes essentially an upgrade from the widely used KEH, as it comes much closer to the angular velocity profile of merger remnants (as seen in a plethora of numerical simulations) and at the same time, it can produce models with an axis ratio between $\sim 0.5$ and $\sim 0.7$ for which the models are quasi-spherical (as defined by the energy density profile in the equatorial plane).

Since the work of \citet{Uryu_etal_2017} there has been a shortage of reference models constructed with the newly proposed rotation laws. The original work of \citeauthor{Uryu_etal_2017} focused on presenting a limited number of selected models. Subsequent studies used the new 4-parameter Uryu+ rotation law for describing rotating strange stars \citep{Zhou_etal_2019}, explored the new 3-parameter Uryu+ rotation law in a Newtonian framework \citep{Passamonti_Andersson_2020}, and evolved in GR a selected model constructed with the new 3-parameter Uryu+ rotation law \citep{Xie_etal_2020}. This was a gap that we tried to partially fill, for the case of the new 4-parameter Uryu+ rotation law, with our present contribution. Expanding the well-tested, robust and fully relativistic, numerical code {\tt rns} with the newly suggested rotation law, provides an additional testing platform, thus supporting the further development of more realistic rotation laws.

Specifically, building on \cite{Uryu_etal_2017}, the implicit treatment of parameters A and B through the ratios $\lambda_1$ and $\lambda_2$, allows one to easily take advantage of information provided from numerical simulations in order to fine-tune the parameter values of the rotation law. For example, according to recent simulations \citep{Hanauske_etal_2017, DePietri_etal_2020}, a value of $\lambda_2 = 1$ could be favoured over $\lambda_2 = 0.5$, for the case of compact remnants from BNS mergers, while $\lambda_1$ falls in a range between 1.7-1.9 for realistic EOS. Furthermore, we also found that using different $\lambda_1$ and $\lambda_2$ values can have an impact on whether the constructed configuration's morphology is quasi-spherical or quasi-toroidal.

Going beyond the characterization of selected equilibrium models, we constructed variants of complete sequence with different $\{\lambda_1, \lambda_2\}$ options, which reflect the different degrees of differential rotation considered. For two of these choices, namely $\{1.5, 1.0\}$ and $\{2.0, 1.0\}$, we were able to find a mass-shedding limit for the new Uryu+ rotation law, which identifies the corresponding sequences as type A solutions, according to the classification in \citet{Ansorg_etal_2009}. The other two choices, namely $\{2.0, 0.5\}$ and $\{1.5, 0.5\}$, result in families of models consisting of both quasi-spheroidal and quasi-toroidal configurations, for which no mass-shedding was encountered. We identified the corresponding sequences as type C solutions, in the \citet{Ansorg_etal_2009} classification.  \citet{Zhou_etal_2019} also explored type C solutions for the case of strange stars with the 4-parameter Uryu+ law \eqref{eq:Uryuetal_rotlaw8}. Here, we present the first classification of \textit{different} types of solutions (A and C) with a rotation law other than the KEH law used in \citet{Ansorg_etal_2009}.

A more thorough exploration of the parameter space for all four parameters $\{p,q,\lambda_1,\lambda_2\}$ of the \citet{Uryu_etal_2017} law will be required to investigate the possible existence of other types of solutions (which are called type B and type D in the classification of \citealt{Ansorg_etal_2009}). 
 
Another result to highlight is the fact that, as shown in Figure~\ref{fig:seqBC_mass_vs_Re_Uryu_vs_KEH}, we systematically obtained smaller radii for our Uryu+ models compared to the corresponding KEH models. For models with $r_p/r_e >0.45$ this can be explained by the fact that, with the Uryu+ law, the angular velocity at the equator, $\Omega_e$ is smaller than for the corresponding models with the KEH law, resulting in a weaker centrifugal force and thus in a smaller radius.   

In Appendix \ref{sec:appendix_CFC}, we investigate the impact of the spatially, conformally flat, spacetime approximation IWM-CFC \citep{Isenberg_2008, Wilson_etal_1996} on the equilibrium structure of models built with the Uryu+ rotation law. We show that down to an axis ratio of $r_p/r_e \sim 0.5$ (which is typical for post-merger remnants), the error introduced by the IWM-CFC approximation remains relatively small. 

We point out, that in some recent simulations of BNS mergers a large effective viscosity was used (as a model for the turbulent viscosity due to the magneto-rotational instability, MRI),  resulting in a rotational profile that quickly becomes uniformly rotating in most parts of the remnant (see \citealt{Shibata_etal_2017, Shibata_Kiuchi_2017, Radice_2017, Fujibayashi_etal_2018}). Whether current simulations have sufficient resolution to accurately inform us about the timescale of  the development of the MRI instability remains to be seen (see \citealt{Kiuchi_etal_2018,Ciolfi_2020,Radice_2020}).

In future work, we aim to expand this initial study from $N=1$ polytropes to tabulated and hot EOS. The inclusion of thermal effects \citep{Kaplan_etal_2014,Camelio_etal_2019} is important to allow for a more realistic reconstruction of post-merger remnants as quasi-equilibrium models, see \citet{Camelio_etal_2020}.
In addition, we also plan to study the dynamics of equilibrium models  built with the Uryu+ differential rotation law, by investigating their oscillation modes and the onset of the low $T/|W|$ instabilities in models (see \citealt{Passamonti_Andersson_2020, Xie_etal_2020}).    

The rotational properties of binary neutron star merger remnants have important astrophysical consequences, affecting e.g. the precise thresholds to prompt or delayed collapse to a black hole (with further consequences on the development of a possible short gamma-ray burst, observable in the electromagnetic spectrum, see e.g. the reviews by \citet{Bauswein_Stergioulas_2019,Friedman_Stergioulas_2020} and references therein). The semi-analytic derivation of an empirical relation for the threshold mass of \citet{Bauswein_Stergioulas_2017} could be updated, using the new rotation law and tabulated EOS. It will also be worth to investigate quasi-universal relations, such as those presented in \cite{Bozzola_etal_2018,Weih_etal_2018}, using the new rotation law.

\section*{Acknowledgements}

We are grateful to the anonymous referee for useful comments that significantly improved the final version of the manuscript. We also thank Gabriele Bozzola and Wolfgang Kastaun for useful discussions and Giovanni Camelio, Tim Dietrich, Stephan Rosswog and Bryn Haskell for advanced sharing of a manuscript on a related topic and for comments on our manuscript. PI gratefully acknowledges support by a Virgo-EGO Scientific Forum (VESF) PhD fellowship. The authors gratefully acknowledge the Italian Istituto Nazionale di Fisica Nucleare (INFN), the French Centre National de la Recherche Scientifique (CNRS) and the Netherlands Organization for Scientific Research, for the construction and operation of the Virgo detector and the creation and support of the EGO consortium.

\section*{Data Availability}

The data underlying this article are available in the article.



\bibliographystyle{mnras}
\bibliography{references}




\appendix

\section{Rescaled equations for code implementation}
\label{sec:appendix_rescalings}

As in all KEH-type codes, so in the {\tt rns} code, key equations and quantities are rescaled with the coordinate equatorial radius $r_e$ for the computations. In this appendix we summarise the relevant rescaled expressions used in this work.

The rescalings for the angular velocity $\Omega$, the gravitationally redshifted angular momentum per unit rest mass $F$, the metric potentials $\rho$, $\gamma$, $\omega$ and the parameters $A$ and $B$, are given by the following expressions
\begin{align}
&\hat{\Omega} = r_e \Omega \;, \quad \hat{F} = F / r_e \;, \quad \hat{\rho} = \rho /{r_e}^2 \;, \quad \hat{\gamma} = \gamma / {r_e}^2, \nonumber \\
&\hat{\omega} = r_e \omega \;, \quad \hat{A} = A /r_e \;, \quad \hat{B} = B / r_e \; .\label{eq:rescalings}
\end{align}
We note that the rescaling for $F$ emerges if we take equation \eqref{eq:F_expr_general} and apply the rescaling relations for $\Omega$, $\omega$ and $\rho$
\begin{equation}
F = u^t u_{\phi} = \frac{r_e \, (\hat{\Omega}- \hat{\omega}) \; s^2 \; (1-\mu^2) \; e^{-2 \, r_e^2 \, \hat{\rho}}}{\left(1-s\right)^2 - \left( \hat{\Omega} - \hat{\omega} \right)^2 s^2 \; (1-\mu^2) \; e^{-2 \, r_e^2 \, \hat{\rho}}} = r_e \hat{F} \; ,\label{eq:F_rescaled_hat}
\end{equation}
where $\mu = \cos \theta$ and $s = r / (r+r_e)$ is the compactified radial coordinate mapping radial infinity to the finite coordinate location $s=1$ \citep{Cook_etal_1992}. Concerning the rescalings of parameters $A$ and $B$, we expect the fractions $\frac{F}{A^2 \Omega_c}$ and $\frac{F}{B^2 \Omega_c}$ from the differential rotation law \eqref{eq:Uryuetal_rotlaw8} to be dimensionless. This demand, together with applying the rescalings for $F$ and $\Omega$, leads to the specific rescalings of parameters $A$ and $B$ reported in \eqref{eq:rescalings}. 

A stepping stone towards obtaining final expressions for a new $\Omega_e$ value and for calculating $\Omega$ everywhere inside the star, is to obtain rescaled relations for $F_e$ and $F_\text{max}$, as well as parameters $A$ and $B$ in the general case, i.e. equations \eqref{eq:Asq_general}, \eqref{eq:Bsq_general}. Therefore, we apply \eqref{eq:F_rescaled_hat} at the equator (where $\mu = 0$ and $s=0.5$) to obtain
\begin{equation}
F_e =  \frac{r_e \, {\upsilon_e}^2}{\left( 1 - {\upsilon_e}^2 \right) \left( \hat{\Omega}_e - \hat{\omega}_e \right)} = \frac{ r_e \left( \hat{\Omega}_e - \hat{\omega}_e \right) e^{-2 \, {r_e}^2 \hat{\rho}_e}}{\left[1 - \left( \hat{\Omega}_e - \hat{\omega}_e \right)^2  e^{-2 \, {r_e}^2 \hat{\rho}_e} \right]} = r_e \hat{F}_e, \label{eq:Fe_hat}
\end{equation}
and at the location of the maximum of $\Omega$ (again at the equatorial plane $ \mu= 0 $) to obtain
\begin{align}
F_\text{max} &= \frac{r_e \, {\upsilon_\text{max}}^2}{\left( 1 - {\upsilon_\text{max}}^2 \right) \left( \hat{\Omega}_\text{max} - \hat{\omega}_\text{max} \right)} \nonumber \\
&= \frac{ r_e \left( \hat{\Omega}_\text{max} - \hat{\omega}_\text{max} \right) {s_\text{max}}^2 \, e^{-2 \, {r_e}^2 \hat{\rho}_\text{max}}}{\left[ \left( 1-{s_\text{max}} \right)^2 - \left( \hat{\Omega}_\text{max} - \hat{\omega}_\text{max} \right)^2 {s_\text{max}}^2 \, e^{-2 \, {r_e}^2 \hat{\rho}_\text{max}} \right]} \nonumber \\
& = r_e \hat{F}_\text{max} \; .\label{eq:Fmax_hat}
\end{align}
Moving on to apply the rescalings \eqref{eq:rescalings} to equations \eqref{eq:Asq_general} and \eqref{eq:Bsq_general}, the general analytic expressions to calculate the parameters $A$ and $B$ are written as
\begin{align}
\hat{A}^2 = \frac{1}{\hat{\Omega}_c} \left[ \left( \hat{F}_e \hat{F}_\text{max} \right)^p  \frac{\left( \lambda_2 \: {\hat{F}_e} ^q - \lambda_1 \: {\hat{F}_\text{max}}^q \right)}{\left[ \: (\lambda_1 - 1) \: {\hat{F}_e}^p - (\lambda_2 -1) \: {\hat{F}_\text{max}}^p \: \right]} \right]^{  \frac{1}{p+q} }, \label{eq:Asq_hat}
\end{align}
\begin{align}
\hat{B}^2 = \frac{\hat{F}_e \hat{F}_\text{max}}{\hat{\Omega}_c} \left[ \frac{\left( \lambda_2 \: {\hat{F}_e} ^q - \lambda_1 \: {\hat{F}_\text{max}}^q \right)}{ \left[ \: \lambda_2 \: (\lambda_1 -1) \: {\hat{F}_e}^{p+q} - \lambda_1 \: (\lambda_2 - 1) \: {\hat{F}_\text{max}}^{p+q} \: \right] } \right]^{1/p} \;. \label{eq:Bsq_hat}
\end{align}
We note that, compared to the initial equations, there are no differences in   $r_e$ terms introduced, i.e. after all rescaled quantities are substituted in both the right and left hand sides, occurrences of $r_e$ eliminate each other.

Next, we turn to the main equations for calculating a new value for $\Omega_e$ and $\Omega$ everywhere inside the star. Substituting the rescalings \eqref{eq:rescalings} into equations \eqref{eq:hydroequil_numsolve} and \eqref{eq:Omega_everywhere_pq} we obtain, respectively
\begin{align}
 {r_e}^2 & \left( \hat{\gamma}_e + \hat{\rho}_e -\hat{\gamma}_p - \hat{\rho}_p \right) +  \ln \left[ 1- \left( \hat{\Omega}_e - \hat{\omega}_e \right)^2   e^{-2 \, {r_e}^2 \hat{\rho}_e}  \right] = \nonumber \\
 &-2 \, \hat{\Omega}_e \frac{\left( \hat{\Omega}_e - \hat{\omega}_e \right) e^{-2 \, {r_e}^2 \hat{\rho}_e}}{\left[ 1 - \left( \hat{\Omega}_e - \hat{\omega}_e \right)^2  e^{-2 \, {r_e}^2 \hat{\rho}_e} \right]} \nonumber \\
&  + \frac{\hat{A}^2 {\hat{\Omega}_e}^2}{2 \, {\lambda_2}^2}  \left\lbrace 2 \, \frac{\hat{A}^2}{\hat{B}^2} \arctan \left( \frac{ {\lambda_2}^2 \left( \hat{\Omega}_e - \hat{\omega}_e \right)^2 e^{-4 \, {r_e}^2 \hat{\rho}_e}}{ \hat{A}^4 {\hat{\Omega}_e}^2 \left[ 1 - \left( \hat{\Omega}_e - \hat{\omega}_e \right)^2 e^{-2 \, {r_e}^2 \hat{\rho}_e} \right]^2} \right) \right. \nonumber \\
& - \sqrt{2} \left[ \arctan\left( 1 - \frac{ \lambda_2 \sqrt{2} \left( \hat{\Omega}_e - \hat{\omega}_e \right) e^{-2 \, {r_e}^2 \hat{\rho}_e}}{ \hat{A}^2 \hat{\Omega}_e\left[ 1 - \left( \hat{\Omega}_e - \hat{\omega}_e \right)^2  e^{-2 \, {r_e}^2 \hat{\rho}_e} \right]} \right) \right. \nonumber \\
& \left. -  \arctan \left( 1 + \frac{ \lambda_2 \sqrt{2} \left( \hat{\Omega}_e - \hat{\omega}_e \right) e^{-2 \, {r_e}^2 \hat{\rho}_e}}{ \hat{A}^2 \hat{\Omega}_e\left[ 1 - \left( \hat{\Omega}_e - \hat{\omega}_e \right)^2  e^{-2 \, {r_e}^2 \hat{\rho}_e} \right]} \right) \right] \nonumber \\
& + \left. \sqrt{2}\tanh^{-1} \left( \frac{ \dfrac{\hat{A}^2 {\hat{\Omega}_e} \sqrt{2}}{\lambda_2}   \dfrac{\left( \hat{\Omega}_e - \hat{\omega}_e \right)  e^{-2 \, {r_e}^2 \hat{\rho}_e}}{\left[ 1 - \left( \hat{\Omega}_e - \hat{\omega}_e \right)^2  e^{-2 \, {r_e}^2 \hat{\rho}_e} \right]}}{\dfrac{\left( \hat{\Omega}_e - \hat{\omega}_e \right)^2  e^{-4 \, {r_e}^2 \hat{\rho}_e}}{\left[ 1 - \left( \hat{\Omega}_e - \hat{\omega}_e \right)^2  e^{-2 \, {r_e}^2 \hat{\rho}_e} \right]^2} + \dfrac{\hat{A}^4 {\hat{\Omega}_e}^2}{{\lambda_2}^2} } \right) \right\rbrace, \label{eq:hydroequil_numsolve_hat}
\end{align}
for the relation to find a new value for $\Omega_e$, and
\begin{equation}
\hat{\Omega} = \hat{\Omega}_c \, \frac{1 + \left( \dfrac{\left( \hat{\Omega} - \hat{\omega} \right) s^2 \, (1 - \mu^2) \, e^{-2 \, {r_e}^2 \hat{\rho}}}{ \hat{B}^2 \hat{\Omega}_c \left[ \left(1 - s \right)^2 - \left( \hat{\Omega} - \hat{\omega} \right)^2 s^2 \, (1 - \mu^2) \, e^{-2 \, {r_e}^2 \hat{\rho}} \right]} \right)^p}{1 + \left[ \dfrac{\left( \hat{\Omega} - \hat{\omega} \right) s^2 \, (1 - \mu^2) \, e^{-2 \, {r_e}^2 \hat{\rho}}}{\hat{A}^2 \hat{\Omega}_c \left[ \left(1 - s \right)^2 - \left( \hat{\Omega} - \hat{\omega} \right)^2 s^2 \, (1 - \mu^2) \, e^{-2 \, {r_e}^2 \hat{\rho}} \right]} \right]^{p+q}}. \label{eq:Omega_everywhere_hat_pq}
\end{equation}
for the relation to calculate a new distribution for $\Omega$ everywhere inside the star.

Finally, applying the rescalings \eqref{eq:rescalings} to the new equation derived for the calculation of the enthalpy \eqref{eq:new_H_distr} and taking into account expression \eqref{eq:Uryu_law_integral} for the integral term, we obtain the following relation for the enthalpy
\begin{align}
H = &H_\text{surface} + \frac{1}{2} \left[ {r_e}^2 \left( \hat{\gamma}_p + \hat{\rho}_p - \hat{\gamma} -\hat{\rho} \right) -\ln \left( 1 - {\upsilon^2} \right) \right] \nonumber \\
& - \hat{F} \hat{\Omega} + \frac{\hat{A}^2 {\hat{\Omega}_c}^2}{4} \left\lbrace \frac{2 \hat{A}^2}{\hat{B}^2} \arctan \left( \dfrac{{\hat{F}}^2 }{ \hat{A}^4 {\hat{\Omega}_c}^2} \right)  \right. \nonumber \\
& - \sqrt{2} \left[ \arctan\left( 1 - \frac{\hat{F} \sqrt{2}}{\hat{A}^2 \hat{\Omega}_c} \right) -  \arctan \left( 1 + \frac{\hat{F} \sqrt{2}}{\hat{A}^2 \hat{\Omega}_c} \right) \right] \nonumber \\
& + \left. \sqrt{2}\tanh^{-1} \left( \frac{ \hat{A}^2 \hat{\Omega}_c \hat{F} \sqrt{2}}{{\hat{F}}^2 + \hat{A}^4 {\hat{\Omega}_c}^2} \right) \right\rbrace . \label{eq:new_H_distr_hat}
\end{align}
We stress that \eqref{eq:hydroequil_numsolve_hat} and \eqref{eq:new_H_distr_hat} are derived for the particular case of $\{p,q\} = \{1,3\}$.

\section{Comparison between GR and IWM-CFC models}
\label{sec:appendix_CFC}

\begingroup
\setlength{\tabcolsep}{4.8pt}
\begin{table*}
        \centering
        \caption{Comparison of the properties of equilibrium models constructed with the Uryu+ rotation law, in full GR and in the IWM-CFC approximation, for  models A9, B10, C5 (which have $r_p/r_e\sim 0.5$).}
        \label{tab:CFC_study}
        \begin{tabular}{ccccccccccccc}
                \hline
                Model & $\epsilon_\text{max}$ & $M$ & $M_0$ & J & $T/|W|$ & $\Omega_c$ & $\Omega_\text{max}$ & $\Omega_e$ & $\Omega_K$ & $R_e$ & $r_e$ & GRV3\\
                & $\left(\times 10^{-3}\right)$ & & & & $\left(\times 10^{-1}\right)$ & $\left(\times 10^{-2}\right)$ & $\left(\times 10^{-2}\right)$ & $\left(\times 10^{-2}\right)$ & $\left(\times 10^{-2}\right)$ & & & \\
                \hline
                \textbf{A9}\\
                GR & 0.40086 & 1.43493 & 1.50612 & 1.96591 & 1.64514 & 1.97549 & 3.95098 & 0.98775 & 2.28329 & 14.5499 & 12.9843 & 6.03066e-05\\
                CFC & 0.40140 & 1.43389 & 1.50509 & 1.96210 & 1.64425 & 1.97906 & 3.95812 & 0.98953 & 2.28836 & 14.5253 & 12.9608 & 1.30367e-02\\
                rel. error (\%) & 0.13321 & 0.07248 & 0.06839 & 0.19380 & 0.05410 & 0.18071 & 0.18071 & 0.18051 & 0.22205 & 0.16907 & 0.18099 \\ \\
                \textbf{B10}\\
                GR & 1.48092 & 2.20535 & 2.41490 & 3.63795 & 1.65695 & 4.10748 & 8.21495 & 2.05374 & 4.15491 & 10.9145 & 8.42730 & 5.64007e-05\\
               CFC & 1.49134 & 2.19289 & 2.40250 & 3.59161 & 1.65022 & 4.15271 & 8.30543 & 2.07636 & 4.20732 & 10.8046 & 8.33003 & 2.81039e-02\\
                rel. error (\%) & 0.70362 & 0.56499 & 0.51348 & 1.27379 & 0.40617 & 1.10116 & 1.10141 & 1.10141 & 1.26140 & 1.00692 & 1.15422 \\ \\
                \textbf{C5}\\
                GR & 3.38103 & 2.16935 & 2.37578 & 3.36450 & 1.64806 & 6.23604 & 12.4721 & 3.11802 & 5.73080 & 8.60458 & 6.10795 & 5.50829e-05\\
                CFC & 3.42815 & 2.14125 & 2.34785 & 3.27299 & 1.63242 & 6.38268 & 12.7654 & 3.19134 & 5.87823 & 8.42468 & 5.95685 & 3.42459e-02\\
                rel. error (\%) & 1.39366 & 1.29532 & 1.17561 & 2.71987 & 0.94899 & 2.35149 & 2.35165 & 2.35149 & 2.57259 & 2.09075 & 2.47383 \\        
                \hline
        \end{tabular}
\end{table*}
\endgroup

\begin{figure*}
    \includegraphics[scale=0.43]{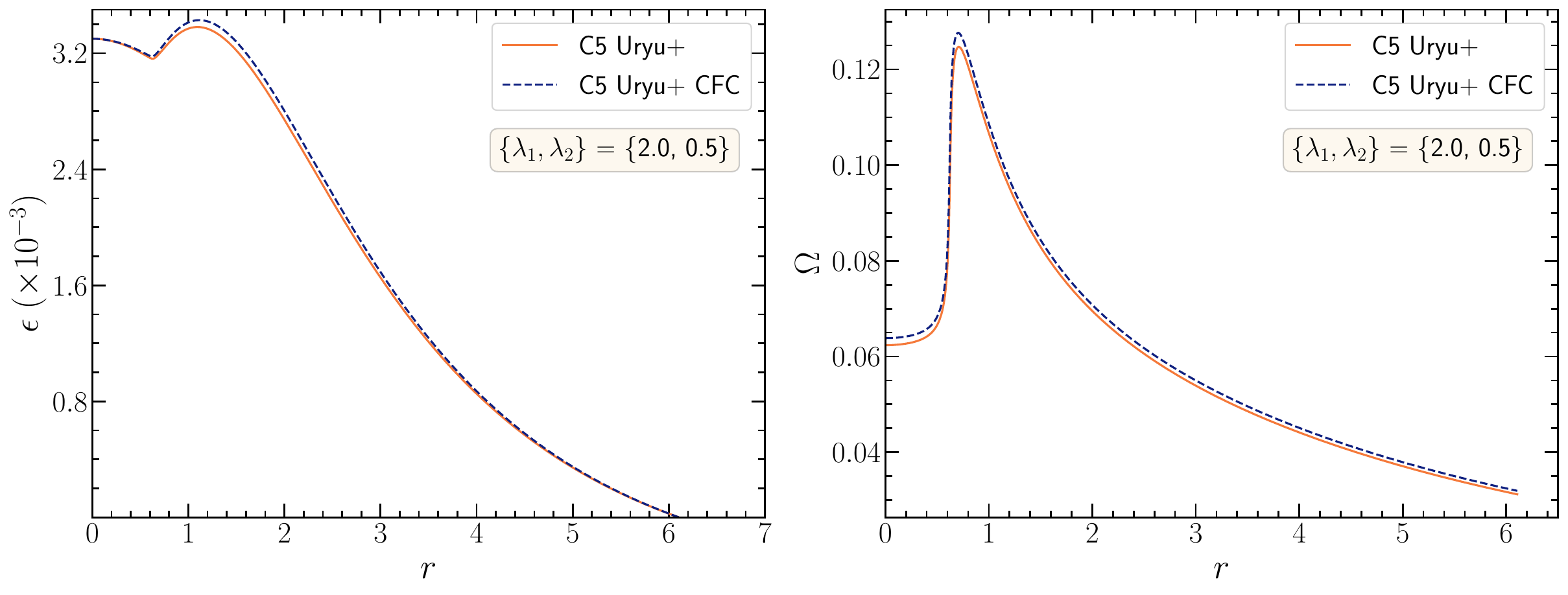}
    \caption{Comparison of the energy density and angular velocity profiles, $\epsilon(r)$ and $\Omega(r)$, versus the coordinate radius $r$, in the equatorial plane for model C5 ($r_p / r_e = 0.5$), calculated in full GR and in the IWM-CFC approximation, for the Uryu+ rotation law.}
    \label{fig:GR_CFC_C5_profiles}
\end{figure*}

Building on our previous work \citep{Iosif_Stergioulas_2014}, we present here a comparison between GR and the spatially, conformally flat, spacetime approximation IWM-CFC \citep{Isenberg_2008, Wilson_etal_1996} for the case of equilibrium models with $r_p/r_e \sim 0.5$, constructed with the Uryu+ differential rotation law \eqref{eq:Uryuetal_rotlaw8}.

In the context of the IWM-CFC approximation, the basic assumption is conformal flatness for the spatial metric 
\begin{equation}
\gamma_{ij} = \psi^4 \eta_{ij}, \label{eq:conformal_flatness}
\end{equation}
where $\psi$ is a conformal factor and $\eta_{ij}$ is the flat metric. Writing the line element in the $3+1$ formalism of general relativity we have
\begin{equation}
ds^2 = -\alpha^2 dt^2 + \gamma_{ij} (dx^i + \beta^i dt)(dx^j + \beta^j dt), \label{eq:line_elem_3plus1}
\end{equation}
where $\alpha$ is the lapse function and $\beta^i$ is the shift vector. If we now consider an axisymmetric star, in spherical-like coordinates and in the absence of meridional circulation, $\beta^{\phi} $ is the only non-zero component of the shift vector $ \beta^{i} $. Therefore, combining equations \eqref{eq:conformal_flatness} and \eqref{eq:line_elem_3plus1}, the line element is expressed as
\begin{equation}
ds^2 = -\alpha^2 dt^2 + \psi^4 (dr^2 + r^2 d \theta^2) +  \psi^4 r^2 \sin ^2 \theta (d \phi + \beta^{\phi} dt)^2  \label{eq:CFC_3+1_metric} \; .
\end{equation}
Going back to the line element \eqref{eq:stationary_axisym_metric} for a stationary, axisymmetric star in full GR and comparing with \eqref{eq:CFC_3+1_metric} above, implies that
\begin{equation}
\alpha = e ^ {(\gamma + \rho) / 2} \; , \qquad
\psi = e ^ {\mu / 2 } = e ^{ (\gamma - \rho) / 4} \; , \qquad 
\beta^{\phi} = -\omega \label{eq:lapse_shift_psi} \; .
\end{equation}
This means that the line element \eqref{eq:stationary_axisym_metric} can take the IWM-CFC form, if
\begin{equation}
\mu = \frac{\gamma - \rho}{2} \label{eq:mu_imposeCFC} \; .
\end{equation}
In summary, a numerical code (such as {\tt rns}) that solves for the full GR metric \eqref{eq:stationary_axisym_metric} can be easily converted to also handle the case of the IWM-CFC approximation, by imposing the condition \eqref{eq:mu_imposeCFC} between the three metric functions instead of solving for the metric potential $\mu$.

Following the above reasoning, we constructed IWM-CFC counterparts for models A9, B10, C5 that exhibit $r_p/r_e \sim 0.5$ assuming the new 4-parameter Uryu+ rotation law \eqref{eq:Uryuetal_rotlaw8} and we compared them to their respective full GR configurations. Our findings are summarised in Table~\ref{tab:CFC_study}. For the lowest density configuration A9, the relative error between GR and CFC is at most $\sim 0.2 \%$ (for the angular velocity of a free particle in circular orbit at the equator $\Omega_K$) and lower for integrated quantities. Moving on to the denser model B10, the relative error is at the $1\%$ level for local quantities, such as the angular velocities and the radii, whereas it remains below $1\%$ for the masses $M$, $M_0$ and the ratio $T/|W|$. This general picture does not change much for the more compact model C5, where the relative error for local quantities rises to $\sim2.5\%$ and remains $\sim1\%$ for the masses and the ratio $T/|W|$. We should point out that the largest relative error between GR and CFC is encountered for the angular momentum $J$ but even that is only $\sim2.7\%$.

As a further check, in \autoref{fig:GR_CFC_C5_profiles} we present the energy density and angular velocity profiles , $\epsilon(r)$ and $\Omega(r)$, for model C5 constructed with the Uryu+ rotation law \eqref{eq:Uryuetal_rotlaw8} both in full GR and in the IWM-CFC approximation. The results underline the robustness of the IWM-CFC method as a viable approximation in numerical simulations of BNS merger remnants, when an accuracy of order $\sim 1\%$ is sufficient.


\section{Additional sequence C model with the KEH rotation law}
\label{seq:extra-C-KEH-model}

In Figures~\ref{fig:seqBC_mass_Uryu_vs_KEH} and \ref{fig:seqBC_mass_vs_Re_Uryu_vs_KEH} we used an additional model along sequence C, constructed with the KEH rotation law that was not part of the original sequence C in \citet{Iosif_Stergioulas_2014}. In Table~\ref{tab:extra-C-KEH-model} we list the detailed properties of this model.

\begingroup
\setlength{\tabcolsep}{3.0pt}
\begin{table*}
        \centering
        \caption{Physical quantities of an additional model along sequence C (with constant  $\epsilon_c = 3.3 \times 10^{-3}$), constructed with the KEH rotation law ($\hat{A}=1$).}
        \label{tab:extra-C-KEH-model}
        \begin{tabular}{lcccccccccccccccc}
                \hline
                 $r_p/r_e$ & $\epsilon_c$ & $\epsilon_\text{max}$ & $M$ & $M_0$ & J & $T/|W|$ & $\Omega_c$ & $\Omega_e$ & $\Omega_K$ & $R_e$ & $r_e$ & GRV3\\
                 & $\left(\times 10^{-3}\right)$ & $\left(\times 10^{-3}\right)$ & &  &  & $\left(\times 10^{-1}\right)$ & $\left(\times 10^{-2}\right)$ & $\left(\times 10^{-2}\right)$ & $\left(\times 10^{-2}\right)$ & & & $\left(\times 10^{-5}\right)$\\
                \hline
                 0.43 & 3.300 & 3.66772 & 2.26784 & 2.48488 & 3.96229 & 1.85476 & 14.6252 & 3.55694 & 5.64641 & 8.81100 & 6.16946 & 4.89800\\
                \hline
        \end{tabular}
\end{table*}
\endgroup

\bsp    
\label{lastpage}
\end{document}